\documentclass[useAMS,usenatbib,onecolumn]{mn2e}
\usepackage[T1]{fontenc}
\usepackage{aecompl}
\usepackage{graphicx}
\usepackage{natbib}
\usepackage{subfloat}
\usepackage{amssymb}
\usepackage{amsbsy}
\usepackage{amsmath}
\usepackage{amsfonts}
\usepackage{color}
\usepackage[normalem]{ulem} 
\usepackage{epsfig}
\usepackage[dvipsnames]{xcolor}

\def\ltsima{$\; \buildrel < \over \sim \;$}
\def\simlt{\lower.5ex\hbox{\ltsima}}
\def\gtsima{$\; \buildrel > \over \sim \;$}
\def\simgt{\lower.5ex\hbox{\gtsima}}

\newcommand{\kjd}[1]{\textcolor{black}{#1}}
\newcommand{\rwqq}[1]{\textcolor{black}{#1}}
\newcommand{\newrw}[1]{\textcolor{black}{#1}}

\newcommand{\rw}[1]{\textcolor{black}{#1}}
\newcommand{\kd}[1]{\textcolor{black}{#1}}
\newcommand{\kdd}[1]{\textcolor{black}{#1}}

\newcommand{\kch}[1]{\textcolor{black}{#1}}

\newcommand{\kref}[1]{\textcolor{black}{#1}}

\title[Constraints on Radial Migration in Spiral Galaxies. II]
{Constraints on Radial Migration in Spiral Galaxies - II. Angular momentum distribution and preferential migration}

\author[Kathryne J. Daniel and Rosemary F. G. Wyse]{Kathryne J. Daniel$^{1,2}$\thanks{E-mail: kjdaniel@brynmawr.edu} and Rosemary F. G. Wyse$^{2,3}$\\
$^{1}$Department of Physics, Bryn Mawr College, Bryn Mawr, PA 19010, USA\\
$^{2}$Department of Physics \& Astronomy, Johns Hopkins University, Baltimore, MD 21218, USA\\
$^{3}$\rw{Institute for Astronomy, University of Edinburgh, Blackford Hill, Edinburgh EH9 3HJ, Scotland, UK}
}

\begin{document}
\date{Accepted January 20, 2018. Received January 3, 2018; in original form October 30, 2017}

\pagerange{\pageref{firstpage}--\pageref{lastpage}} \pubyear{2018}

\maketitle

\label{firstpage}

\begin{abstract}

The orbital angular momentum of individual stars in galactic discs can be permanently changed through torques from transient spiral patterns. Interactions at the corotation resonance dominate these changes and have the further property of conserving orbital circularity. We derived in an earlier paper an analytic criterion that an unperturbed stellar orbit must satisfy in order for such an interaction to occur i.e.~for it to be in a trapped orbit around corotation. We here use this criterion in an investigation of how the efficiency of induced radial migration for a population of disc stars varies with the angular momentum distribution of that population. We frame our results in terms of the velocity dispersion of the population, this being an easier observable than is the angular momentum distribution. Specifically, we investigate how the fraction of stars in trapped orbits at corotation varies with the velocity dispersion of the population, for a system with an assumed flat rotation curve. Our analytic results agree with the finding from simulations that radial migration is less effective in populations with \lq hotter' kinematics. We further quantify the dependence of this trapped fraction on the strength of the spiral pattern, finding a higher trapped fraction for higher amplitude perturbations.
 
\end{abstract}
\begin{keywords}
galaxies: evolution, galaxies: kinematics and dynamics, galaxies: structure
\end{keywords}

\section{Introduction}\label{sec:Introduction}

Internal, secular mechanisms can have a significant impact on the
evolution of galactic \kch{disc}s \citep[see~e.g.][]{Sellwood14}.
\rw{For example}, it is now recognized that \kch{disc} stars can migrate
\rw{radially, moving} of order a \kch{disc} \kch{scale-length} over their \rw{lifetime,}
through successive resonant interactions with transient gravitational
perturbations (\rw{for observational evidence,} see
e.g.,~\kd{\cite{WFD96,Bovy12,RS12,Hayden15MNRAS,Kordopatis15MNRAS,Loebman16}}
 and \rw{for theoretical investigations see e.g.~}\kd{\cite{SB02,Roskar08b,SB09b,Loebman11,Minchev12,RDL13,VCdON16,MartinezMedina17,PAR17,SM17}}). \cite{LBK72} established that stars
on close to circular orbits can exchange orbital angular momentum with
steady (or slowly growing) spiral waves at the major resonances,
namely corotation and \kch{the} inner/outer Lindblad resonances.
These authors further demonstrated that the change in angular momentum
at corotation occurs without any increase in \kch{the energy
associated with non-circular motion}, whereas torques at the Lindblad
resonances do \kch{increase orbital} eccentricity \rw{(i.e.~cause kinematic heating of the population}).

\cite{SB02}, in an influential paper, showed that \rw{for} a star in a \lq\lq
trapped orbit"\footnote{We use the term \lq\lq trapped orbit" for
members of the \lq\lq horseshoe" orbital family associated with the
corotation resonance, in order to emphasise their role as the first
stage in radial \kch{migration}.} around the corotation radius of a
\rw{\it{transient\/}} spiral \rw{pattern, the angular momentum change can be permanent, leading to radial migration of the star}.  \rw{While the spiral pattern exists, a trapped star is
repeatedly torqued, \kd{causing its 
mean orbital radius (hereafter denoted $R_L$, under the assumption that it  equals the guiding
centre radius in the epicyclic approximation, i.e.~the radius of a
circular orbit with angular momentum equal to that of the star at that instant) to
move outwards then inwards across corotation. The star's guiding centre radius oscillates in} time until the torques cease when the
transient spiral perturbation dies away, restoring axisymmetry and leaving the star with \kd{a} new \kd{value for}  orbital angular momentum and guiding radius} (except in the unlikely case that
it is dropped back where it started). \rw{There is}  no associated increase in
\kch{non-circular motion}. \cite{SB02} showed that interactions at the
corotation resonance \kch{dominate} the angular momentum changes
\kch{in the disc} (compared to interactions at the Lindblad
resonances)\kch{. W}e will refer to the changes in mean orbital radius
due to the corotation interactions as \lq\lq radial \kch{migration}''.

All orbits have an associated circular frequency and so, provided
there exists an appropriately matched spiral \rw{(with pattern speed approximately equal to the stars's circular frequency)} for a long enough period
of time, any star could in principle migrate radially.  However, as
pointed out by \cite{SB02}, stars on eccentric orbits cannot \lq\lq
keep station" with the spiral pattern at the radius of corotation and
therefore radial \kch{migration} is expected to be less important for
kinematically \rw{hotter} stellar populations.  Our investigations in this
paper are aimed at quantifying this expectation.

\begin{figure}
\begin{center}
\includegraphics[scale=1]{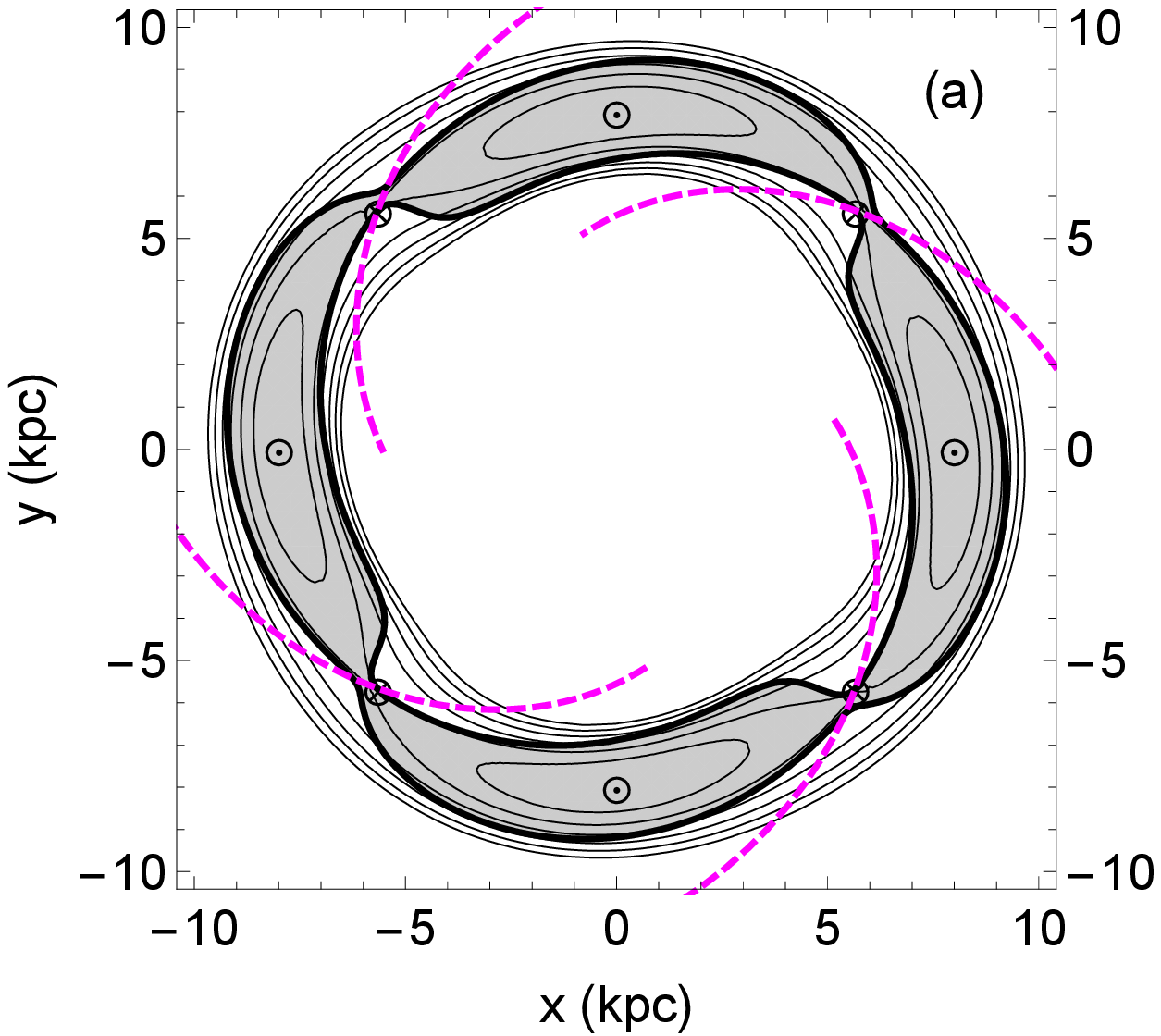}\\
\includegraphics[scale=0.5]{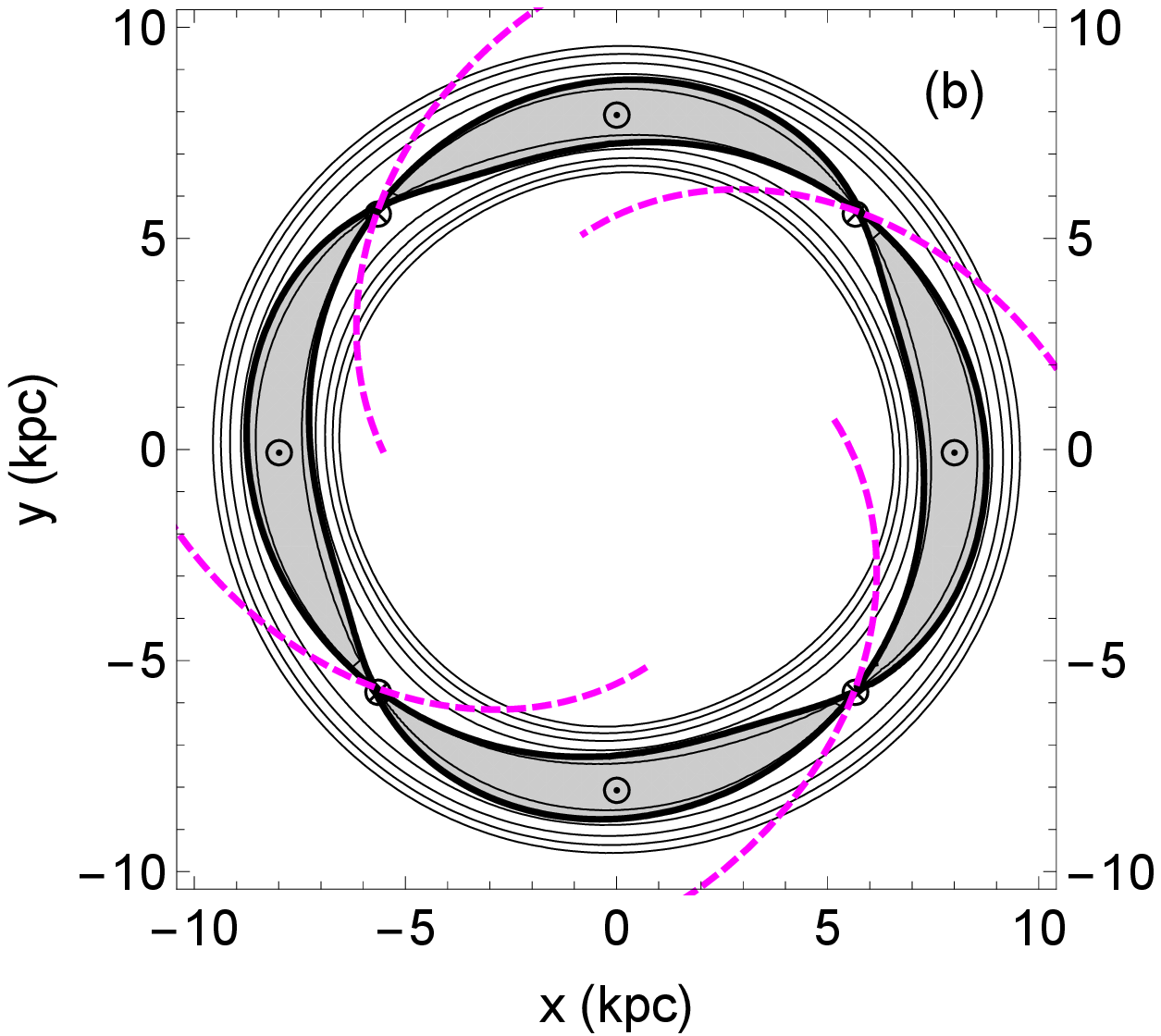}
\includegraphics[scale=0.5]{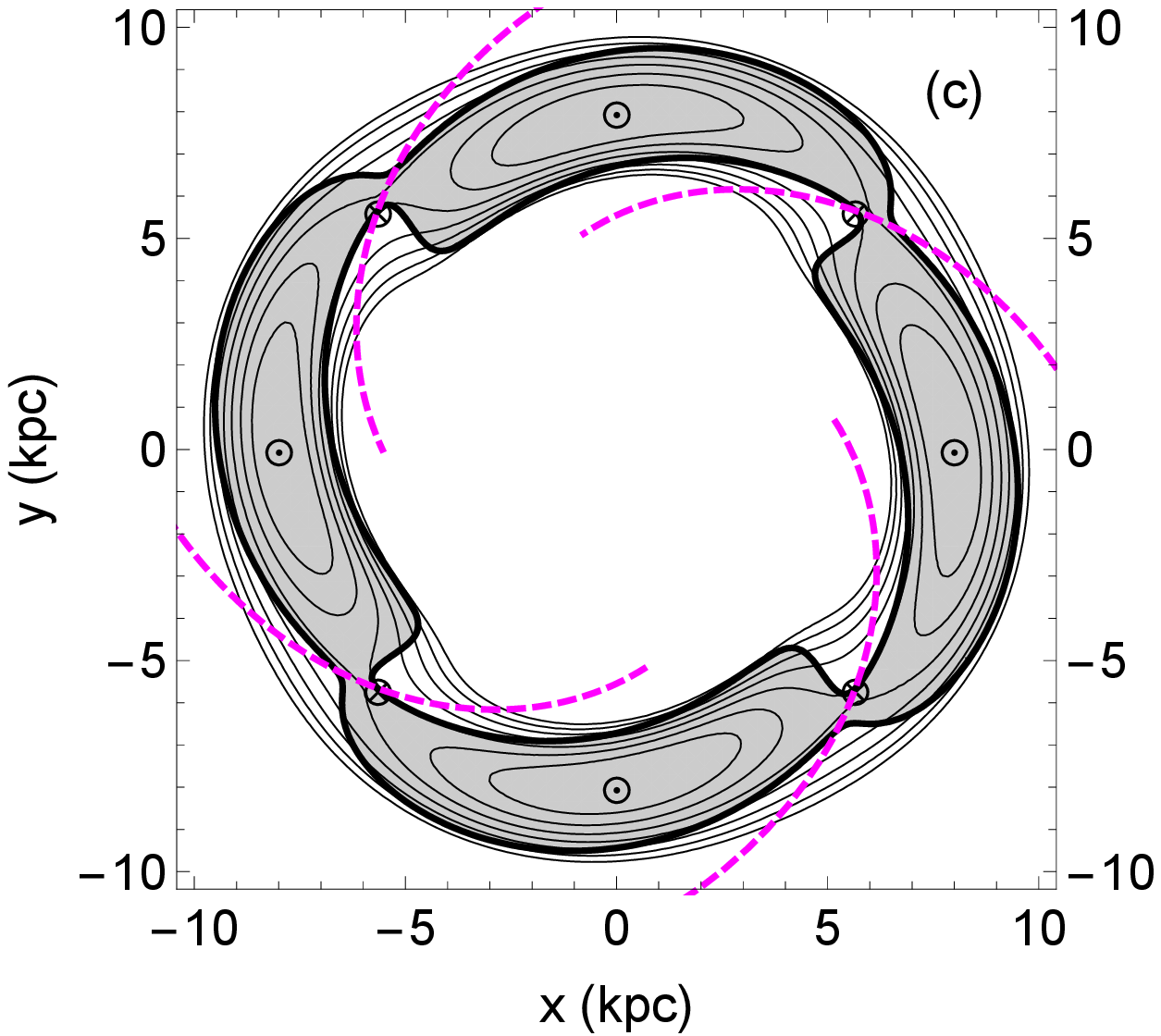}

\caption{Contours of the effective potential, $\Phi_{eff}$, for a
trailing spiral pattern with \rw{fixed parameter values given in Table~\ref{tbl:Values}, radius of corotation at $R_{CR}=8$~kpc, and 
 fractional amplitude for the spiral
surface density} equal to $\epsilon_\Sigma=0.3$~in~panel~(a),
$\epsilon_\Sigma=0.1$~in~panel~(b), and $\epsilon_\Sigma=0.5$~in~panel~(c).
\kch{Contours are shown in $400$~km$^2$~s$^{-2}$ intervals around the
value of $\Phi_{eff}$ at \rw{the peak amplitude of the spiral pattern} and are within
2~kpc of corotation.}  The \rw{locations of the} peaks of the spiral perturbation are shown
as thick, dashed (magenta) curves.  The local maxima in $\Phi_{eff}$
(located between the spiral arms) are marked with the symbol $\odot$
and the saddle points (the deepest part of the spiral potential at
$R=R_{CR}$) are marked with $\otimes$ \rw{(note these are on the magenta dashed curves marking the maxima of the spiral pattern}).  The capture region has a thick
(black) outline and is shaded grey.  The model \rw{shown in panel~(a) is adopted} for much of the analysis in this paper.  }
\label{fig:RMEII_CaptureRegion}
\end{center}
\end{figure}

We \rw{earlier} \citep[hereafter Paper~I]{DW15} derived an analytic
expression for a \lq\lq capture criterion" that determines whether or
not a \kch{\kch{disc}} star with given instantaneous
\kch{4D}~phase-space coordinates \rw{may be}  in a trapped orbit around the
corotation radius of an imposed spiral pattern (see
Appendix~\ref{sec:Capture} for a brief review). The physical
\rw{interpretation} of this capture criterion (\rw{given in} equation~\ref{eqn:Lambda_nc2})
is that the orbital angular momentum of a star is the most important
quantity that determines whether or not that star is in a trapped
orbit.  Indeed, the criterion may be re-stated \kd{for kinematically warmed stellar orbits, where such a}  star \rw{will be} in
a trapped orbit provided \rw{the instantaneous value of its} guiding centre radius lies
within a \rw{certain region, the} \lq\lq capture region'' surrounding the corotation
radius. This \rw{capture} region is \kch{approximately delineated by} the closed
contours of the effective potential ($\Phi_{eff}$, the potential in
the frame that rotates with the spiral pattern) \citep[see][ \S2.2.1
\kch{for a \rw{more} complete description}]{DW15}.  The capture regions for
three different strengths of spiral perturbation are illustrated in
Figure~\ref{fig:RMEII_CaptureRegion}, as shaded areas outlined with a
thick, black line (see Table~\ref{tbl:Values} for the values of the
fixed parameters \rw{of the model}; the radius of corotation is \rw{here} chosen to be \rw{at} 
$R_{CR}=8$~kpc).  Contours of the effective potential are shown by
thin lines while the local maxima in the surface density of the spiral
pattern are indicated by thick, dashed (magenta) curves.  The \rw{maximum} radial
extent of the capture region equals the maximum possible change in
guiding center radius - and hence \rw{distance of} induced radial \kch{migration} -
from the given (transient) spiral pattern. This maximum \rw{width equals} a few
kiloparsec, of order the \kch{disc} \kch{scale-length}, for the
parameter values \rw{chosen for} Figure~\ref{fig:RMEII_CaptureRegion}.  Most
stars will migrate \rw{significantly} less than the maximum distance \kd{\citep[\rw{
quantified} in ][]{DW18}.}

Circular orbits do not exist in a non-axisymmetric potential due to
non-conservation of angular momentum. We can however consider a star
moving on an orbit along which, at all times, the (time-dependent)
spatial coordinates of \kch{the} guiding centre equal those of the
star \kch{(its orbit has zero random energy)}. In this \rw{case, should the star
be in a trapped orbit, it is} physically located within the capture region.  A
star \rw{on an orbit with finite random  energy}  meets the capture criterion and
is in a trapped orbit \rw{provided} its \textit{guiding \rw{centre}} is
within the capture region, even while the star's spatial coordinates
place it outside the capture region.  The typical amplitude of radial
epicyclic excursions for a population of \kch{stars with} radial
velocity dispersion $\sigma_R$, moving in a potential with local
epicyclic frequency $\kappa$, may be approximated by the ratio $
\sigma_R / \kappa$. Stars in the local Galactic disc with $\sigma_R
\sim 30\,\mathrm{km\,s^{-1}}$ have excursions of order 1~kpc. \rw{Taking the finite width of the
capture region into account, for the model of panel (a) Figure~\ref{fig:RMEII_CaptureRegion}, stars that are members of such a population physically} located $\simgt 2$~kpc from
corotation could meet the capture criterion \rw{(guiding centre within the capture region)}. Further, for a flat
rotation curve the epicyclic frequency decreases with radial
coordinate as $\kappa \propto R^{-1}$ so that for given value of
velocity dispersion, the epicyclic excursions are larger in the outer
disc and stars located further from corotation can meet the capture
criterion.

\rw{The radial period of trapped orbits for a given spiral
perturbation is longer for larger excursions.\footnote{\kd{Simple harmonic motion is not a good approximation for trapped orbits \citep[see, for example,][]{Contopoulos73}.}}
Our numerical experiments} show that the largest
trapped orbits have periods a few times longer than a galactic year at
corotation.  A trapped orbit that has guiding centre excursions of
order $\sim 1$~kpc from corotation, in the potential illustrated by
Figure~\ref{fig:RMEII_CaptureRegion}, panel(a), takes $\sim 8\times 10^8$~years
to complete a full \rw{radial} oscillation, whereas a galactic year at
corotation is $\sim 2.2\times 10^8$~years and the epicyclic period is
$\sim 1.6 \times 10^8$~years.  The \rw{radial} period of trapped
orbits with the largest excursions places a lower limit on the
timescale for full phase mixing \kd{of trapped orbits around corotation}. 
Therefore the minimum \rw{lifetime for a transient spiral that can
induce radial migration over a significant distance} is a few times a
galactic year at corotation.

In this paper, we apply the capture criterion to a series of model
\kch{disc} galaxies with the aim \rw{of determining} the (initial)
distribution and fraction of a population of stars in trapped orbits
\rw{due to an imposed spiral pattern}.\footnote{We do not \rw{include possible 
additional} perturbations to the gravitational potential, such as a
bar or Giant Molecular Clouds, that could scatter stars out of
trapped orbits associated with the spiral of interest.}  While orbital
angular momentum is the physical parameter that is most important in
determining whether or not a star meets the capture criterion, the
orbital angular momentum distribution is far from trivial to determine
from observations. We therefore use velocity dispersion as a proxy.
Our analysis aims to quantify how the velocity dispersion of a
population \rw{determines} the fraction of \kch{disc} stars that satisfy
the capture criterion.

This paper is organized as follows: we outline our approach in
\S\ref{sec:Approach}, with the adopted \rw{two-dimensional disc} potentials described
in \S\ref{sec:Model}, the analytic forms of the stellar phase-space
distribution functions in \S\ref{sec:DF}, and derive the resulting
distribution of orbital angular momentum for each model in
\S\ref{sec:AMDistribution}.  We apply our capture criterion to
\kch{simple versions of} these models in \S\ref{sec:CapturedFraction}
\kch{in order to} focus on how the trapped population depends on
\rw{both invariant and radially varying forms of  the stellar} velocity
dispersion in \S\ref{sec:CapturedFractionSigNotRadDep} and
\S\ref{sec:CapturedFractionSigRadDep}, respectively.  We investigate
how the trapped population depends on other parameters of the model
(such as strength of the assumed spiral) in
\S\ref{sec:FncDepOfTrapFrac}.  \kch{We apply our capture criterion to
the full models of the Galactic disc in \S\ref{sec:CapFracInModels}.}
\S\ref{sec:Discussion} places our results in the context of earlier
investigations and \S\ref{sec:Conclusions} summarises our conclusions.
The Appendix contains a short discussion of the salient points in the
derivation of the analytic capture criterion from Paper~I.

\section{Approach}\label{sec:Approach}

This section describes our models for the potential of the \kch{disc}\kch{, the stellar kinematics,} and \rw{the} spatial distribution of \kch{stars}.  We will consider only motion in the equatorial plane of an underlying axisymmetric potential (plus imposed spiral perturbation). Unless otherwise stated, we use the values given in Table~\ref{tbl:Values} when quantitative calculations are required.

\begin{table}
\begin{center}
\caption{Adopted parameter values when a quantitative analysis is required, unless otherwise stated.}
\begin{tabular}{lcc}
\hline
Parameter & Symbol & Value \\
\hline
Circular velocity (assumed constant with radius)& $v_c$ & 220~km~s$^{-1}$ \\
Characteristic scale of underlying potential & $R_p$ & 1~kpc \\
Number of spiral arms & $m$ & 4 \\
Spiral pitch angle & $\theta$ & $25^\circ$ \\
Fractional amplitude of spiral surface density & $\epsilon_\Sigma$ & 0.3 \\
\kd{Disc} surface density normalisation & $\Sigma(R=8$~kpc$)$ & 50~M$_\odot$~pc$^{-2}$ \\
Scale length for surface density & $R_d$ & 2.5~kpc \\
Radial width of an annulus & \rw{$\delta R$} & 0.1~kpc \\
\hline
\end{tabular}
\label{tbl:Values}
\end{center}
\end{table}

\subsection{Model for the \kch{Disc} Potential}\label{sec:Model}

{We assume that the potential governing the motion within an
infinitely thin disc, $\Phi(R,\phi)$ in cylindrical coordinates,
results} from an underlying axisymmetric potential, $\Phi_0(R)$, plus
a spiral perturbation {that is time-dependent in the inertial frame}, $\Phi_1(R,\phi,t)$.  {We adopt the logarithmic} axisymmetric potential
that gives rise to a flat rotation curve,
\begin{equation}\label{eqn:AxisymmetricPotential}
\Phi_0(R) = v_c^2 \ln(R/R_p) ,
\end{equation}
where $v_c$ is the {(constant)} circular velocity and $R_p$ is the characteristic {scale of} the potential.  We superpose an $m$-armed spiral density wave\footnote{{We assume a density wave for two reasons: its form is analytic and the pattern speed is
invariant with radius.}} \citep{LS64,LYS69,BT08} {of the form} 
\begin{equation}\label{eqn:SpiralPotential}
\Phi_1(R,\phi) = |\Phi_s(R)| \cos [m\,\cot\theta \ln(R/R_{CR}) - m\,\phi]
\end{equation}
in the {(non-inertial)} frame {rotating}  with the spiral pattern {speed ($\Omega_p$)}.  {$R_{CR}=v_c/\Omega_p$ is the radius of corotation and} $\theta$ is the pitch angle of the spiral arm (measured counter-clockwise from a line of constant azimuth).  The {zero of the azimuthal coordinate, $\phi$, \rw{may be chosen arbitrarily and for convenience we chose it to be at the maximum of $\Phi_{eff}$} aligned with}  the positive $x$-axis in Figure~\ref{fig:RMEII_CaptureRegion}. Further, {$\phi$} increases in the counter-clockwise direction. 

The amplitude of the spiral potential \rw{of equation~\ref{eqn:SpiralPotential}} is given by
\begin{equation}\label{eqn:SpiralAmplitude}
\Phi_s(R)=\dfrac{2 \pi G \Sigma(R) \epsilon_\Sigma}{k(R)},
\end{equation}
where $\Sigma(R)$ is the surface density of the \kch{disc}, $\epsilon_\Sigma$ is the fractional amplitude in surface density of the spiral pattern, and $k(R)$ is the radial {wavenumber of the spiral pattern, defined as:}
\begin{equation}\label{eqn:WaveNumber}
k(R) = \dfrac{m}{R \tan\theta}.
\end{equation}
This {form}  for the {spiral} potential has a radially dependent {amplitude} ($\Phi_s(R)$) for any given fractional {enhancement} in surface density {($\epsilon_\Sigma$)}.   {Further, the radius of corotation enters explicitly (see equation~\ref{eqn:SpiralPotential}); 
we return to these points in 
\S\ref{sec:FncDepOfTrapFrac} below.}  

\subsection{Models for the Kinematics of the \rw{Axisymmetric Disc}}\label{sec:DF}

{We adopt two different prescriptions for the stellar phase space distribution function, $f$, to facilitate exploration of the role of the angular momentum distribution. The first assumes simple Gaussians for each component of motion, while the second, taken from \citet{Dehnen99b}, produces an asymmetric azimuthal-velocity distribution, with a tail to lower values (matching observations in the local disc). We \kch{assume} that the phase space distribution is unchanged during the turn-on of the spiral pattern \kd{(discussed in the introduction to \S\ref{sec:CapturedFraction})}.}
 
\subsubsection{Gaussian Velocity Distribution}\label{sec:fG}

The 4D phase space distribution function for { 2D motion in the \kch{disc} plane can be written in} the form \citep[the 6D form is given by][equation~4.22]{BT08},
\begin{equation}
f(\mathbf{x},\mathbf{v}) = \Sigma(\mathbf{x}) \,P(\mathbf{v})_\mathbf{x},
\end{equation}
where $P(\mathbf{v})_\mathbf{x}$ is the velocity distribution function at coordinate $\mathbf{x}$ and $\Sigma(\mathbf{x})$ is the surface density of the \kch{disc}. \rw{We assume that the disc follows an exponential surface 
density profile} \citep[e.g.,][]{Freeman70}:
\begin{equation}\label{eqn:SurfaceDensityProfile}
\Sigma(R)=\Sigma_0 e^{-R/R_d},
\end{equation}
with radial scale length $R_d$.

\rw{The velocity distribution function may be approximated by  a 2D Gaussian, which we will denote by $P_G(\mathbf{v})_\mathbf{x}$.  This form cannot reproduce the observed non-Gaussian skew in the azimuthal velocities at a given location in the \kd{disc}, so  in our models below that adopt the Gaussian, asymmetric drift is incorporated by a simple offset in the mean.}

{We decompose the \rw{velocity} (in the inertial frame) into circular velocity plus random velocity:} 
\begin{equation}\label{eqn:vran}
\begin{array}{cl}
\mathbf{v} &= v_R\,\mathbf{\hat{R}}+v_\phi\,\boldsymbol{\hat{\phi}}\\
&=v_{ran,R}\,\mathbf{\hat{R}}+(v_c+v_{ran,\phi})\,\boldsymbol{\hat{\phi}}\\
&=\mathbf{v}_{ran}+\mathbf{v}_c.
\end{array}
\end{equation} 
 
{The distribution of the radial component of the (random)
velocity is centred on $\langle v_{ran,R}\rangle=0$, with standard
deviation equal to the assumed radial velocity dispersion,
$\sigma_R(\mathbf{x})$.}  The azimuthal velocity distribution is
centred at {$v_c+\langle v_{ran,\phi}\rangle \equiv v_c-v_a$,
where $v_a$ is the asymmetric drift. As discussed in \citet{BT08} (their sections 4.4.3 and 4.8.2a), for any distribution function that represents a kinematically cool \kch{disc} (i.e.~the random motions are much less than the circular velocity), the asymmetric drift velocity scales approximately as the square of the radial velocity dispersion. Our models are intended to resemble the Milky Way \kch{disc} and we adopt the following relation, consistent with local observations  and with the \rw{(simplified)} radial Jeans equation evaluated for the solar location \citep{DB98}}:
\begin{equation}\label{eqn:AD}
v_a= \sigma_R^2/C, 
\end{equation} 
where the scaling constant is set to $C=80\,\mathrm{km\,s}^{-1}$.

\rw{We further simplify the analysis by assuming} a fixed ratio of
radial to azimuthal velocity dispersions, equal to that obtained for
the local disc, $\sigma_R/\sigma_\phi=0.63$ \citep{Nordstrom04}
(\rw{which is} very similar to the value of $1/\sqrt{2}$ expected for
a flat rotation curve; \citet[][equation~3.100]{BT08}).  {This allows
us to express the 2D probability distribution for velocity at a given
spatial coordinate in terms of the radial \kch{velocity} dispersion
alone, as:}

\begin{equation}\label{eqn:fG}
\begin{array}{cl}
P_G(\mathbf{v})_\mathbf{x}
& =\dfrac{1}{2 \pi \sigma_R \sigma_\phi}\exp\left\lbrace-\dfrac{(v_R-\langle v_R\rangle)^2}{2 \sigma_R^2}-\dfrac{(v_\phi-\langle v_\phi\rangle)^2}{2 \sigma_\phi^2}\right\rbrace\\
& = \dfrac{0.63}{2 \pi \sigma_R^2(R)}\exp\left\lbrace-\dfrac{v_{ran,R}^2}{2 \sigma_R^2(R)}-\dfrac{\kref{(v_{ran,\phi}+\sigma_R^2(R)/C)^2}}{2 (\sigma_R(R)/0.63)^2}\right\rbrace, 
\end{array}
\end{equation}

\noindent where we have explicitly indicated that  the radial velocity dispersion may vary with radial coordinate. \kref{The positive sign in front of the asymmetric drift term of the exponential in the second row of equation~\ref{eqn:fG} sets the mean azimuthal velocity to slower rotation for kinematically hotter populations}.  Populations with different \rw{prescriptions} for $\sigma_R$ will also have different orbital angular momentum distributions (and hence different fractions of stars in trapped orbits, for given imposed spiral pattern). \rw{This is discussed}  in more detail in \S\ref{sec:AMDistribution} below.

The full Gaussian distribution function we adopt in \rw{two of the models investigated below}  thus has the form
\begin{equation}\label{eqn:fGfull}
\begin{array}{cl}
f_G(\mathbf{x},\mathbf{v}) &= \Sigma(\mathbf{x}) \,P_G(\mathbf{v})_\mathbf{x}\\
&=\Sigma_0 e^{-R/R_d} \,\dfrac{0.63}{2 \pi \sigma_R^2(R)}\exp\left\lbrace-\dfrac{v_{ran,R}^2}{2 \sigma_R^2(R)}-\dfrac{\kref{(v_{ran,\phi}+\sigma_R^2(R)/C)^2}}{2 (\sigma_R(R)/0.63)^2}\right\rbrace.
\end{array}
\end{equation}

\subsubsection{Distribution Function for a Warmed \kch{Disc}}\label{sec:fNew}

{The distribution function for \rw{an axisymmetric}  stellar \kch{disc} with finite
random motions may be obtained by \kch{\lq warming\rq} the distribution function
for an initially cold \kch{disc} (where all stars are on circular
obits). This approach was pioneered by \cite{Shu69} and revisited by
\cite{Dehnen99b}. In these warmed-\kch{disc} 
distribution functions, the radial excursions of
stars on non-circular orbits from the inner \kch{disc} naturally give rise
to asymmetric drift and a low velocity tail in the azimuthal velocity
distribution \citep{Shu69}. There are several choices for the form of the warmed-\kch{disc} distribution function and \rw{we  adopt} that given as $f_{new}$
in \citet[][his equation (10)]{Dehnen99b}, which we denote $f_D$: }
\begin{equation}\label{eqn:fS}
f_D(E,L_z) = \dfrac{\Sigma(R_E)}{\sqrt{2}\pi \sigma_R^2(R_E)} \exp\left\lbrace\dfrac{\Omega(R_E)[L_z-L_c(R_E)]}{\sigma_R^2(R_E)}\right\rbrace,
\end{equation}

\noindent \kd{where} \rw{$E$ is orbital energy, which enters the right-handside through the parameter $R_E$ which is the orbital radius for a star in a circular orbit with energy $E$, $\Omega(R)$ is the circular frequency at a given radial coordinate, $L_z$ represents orbital angular momentum {(about the $z$-axis)} and $L_c(R)$ is the orbital angular {momentum of} a star in a circular orbit at radius $R$.  The surface density generated from $f_D$ is well approximated by an exponential at radii greater than a \kch{disc} scale length \citep{Dehnen99b}.}


\subsubsection{Prescriptions for the Radial Velocity Dispersion Profile}\label{sec:SigRNormalization}

We explore two different types of prescription for the radial velocity
dispersion profile, with two different aims. \rw{The first, adopted in
the analyses of sections \S\ref{sec:CapturedFractionSigNotRadDep}
\kch{and \S\ref{sec:FncDepOfTrapFrac}}, assumes a constant velocity
dispersion, independent of radial coordinate.}  This assumption allows
us to \kch{estimate how the kinematic temperature} affects the
fraction of stars in trapped orbits \rw{(after imposition of a spiral
pattern)}, \kch{and particularly to} isolate and evaluate \kch{the
effects of} asymmetric drift \rw{(equation~\ref{eqn:AD}). Since
stellar velocity dispersions are observed to vary with radius, this
first approach is valid only when applied over a limited radial
range. This leads to our second approach, adopting different forms of
the possible variation of velocity dispersion with radius.  We
investigate the results from three specific radial velocity dispersion
profiles (detailed in Table~\ref{tbl:Models}) in sections}
\S\ref{sec:CapturedFractionSigRadDep} \kch{and
\S\ref{sec:CapFracInModels}}.

\begin{table}
\begin{center}
\caption{Important parameters for the models presented in \S\ref{sec:CapturedFractionSigRadDep}.  
}
\begin{tabular}{lclcc}
\hline
Model Name & $DF$ & $\sigma_R(R)$ & $R_\sigma$ & Normalisation Value\\
 & & & & $\sigma_R(R=8$~kpc) \\
\hline
Model~$\Sigma$ & $f_G$ & \kch{$\propto\Sigma^{1/2}$} & $2R_d$ & \kch{$35$~km~s$^{-1}$}\\
Model~Q & $f_G$ & \kch{$\propto R\,e^{-R/R_\sigma}$} & $R_d$ & \kch{$28$~km~s$^{-1}$}\\
Model~W & $f_D$ & \kch{$\propto e^{-R/R_\sigma}$} & $3R_d$ & \kch{$35$~km~s$^{-1}$}\\
\hline
\end{tabular}
\label{tbl:Models}
\end{center}
\end{table}

\rw{Table~\ref{tbl:Models} also gives the form of the adopted velocity
distribution function for three fiducial models (denoted $\Sigma$, Q
and W).  Model~$\Sigma$ incorporates the form of  $\sigma_R(R)$ that is appropriate for a  locally isothermal,
self gravitating \kch{disc} with scale height independent of radius
and with fixed ratio of $\sigma_R/\sigma_z$, namely} 
$\sigma_R^2(R)\propto\Sigma(R)$ \citep{vdKS81,LF89}.  \rw{For  an assumed} exponential surface density profile, 
$\sigma_R(R)=\sigma_{R,0} e^{-R/ R_\sigma}$, where $R_\sigma=2 
R_d$. \rw{As noted in Table~\ref{tbl:Models}, the radial velocity dispersion is
normalised such that $\sigma_R(R=8$~kpc$)=\kch{35}$~km~s$^{-1}$,} in
order to approximate solar neighbourhood values.  In this model \rw{(Model $\Sigma$)}, we
adopt a Gaussian distribution function, $f_G$, which includes the
prescription for asymmetric drift given by equation~\ref{eqn:AD}.

\rw{Model~Q} also uses \rw{the} Gaussian velocity distribution \rw{function} ($f_G$) and \rw{the} prescription for asymmetric drift (equation~\ref{eqn:AD}). \rw{The velocity dispersion profile, however, is determined by setting} the Toomre Q parameter \citep{Toomre64} for local stellar \kch{disc} stability to 1.5 at all radii,
\begin{equation}\label{eqn:Q}
Q\equiv \dfrac{\sigma_R(R)\, \kappa(R)}{3.36\, G\, \Sigma(R)}=1.5. 
\end{equation}

\rw{Such a kinematically warm \kch{disc} is locally stable but may
support global instabilities at all radii
\citep[see][]{Sellwood14}. We are interested in modelling a \kch{disc}
that could have transient instabilities at any radius.  \cite{SC14}
found that transient \textit{modes} can occur in their N-body
simulations even as the \kch{disc} is heated to $Q>1.5$ by spiral
activity.}  In a \kch{disc} with a flat rotation curve,
$\kappa(R)\propto R^{-1}$ \rw{so that  $\sigma_R(R) \propto R\,
e^{-R/R_d}$ for constant $Q$. We normalise the model by setting  the surface density at $R = 8$~kpc to equal 50~$M_\odot$~pc$^{-2}$ (Table~\ref{tbl:Values}) in order to approximate  the local disc \citep{KG91}}.

\rw{Model~W is constructed} under the assumption of a kinematically warmed \kch{disc} using \rw{the  modified Shu distribution function $f_D$ introduced in section~\ref{sec:fNew}}.
\cite{Dehnen99b} \rw{found} that the radial velocity dispersion profile \rw{derived from}  the second moments of $f_D$ \rw{is} well approximated by an exponential radial profile \rw{such that} $\sigma_R(R)=\sigma_{R,0} e^{-R/ R_\sigma}$, \rw{with}  $R_\sigma=3 R_d$. \rw{Unless \kd{otherwise} noted,} the value of $\sigma_{R,0}$ is again normalised such that the radial velocity dispersion $\sigma_R(R=8$~kpc$)=\kch{35}$~km~s$^{-1}$.   

\begin{figure}
\begin{center}
\includegraphics[scale=1]{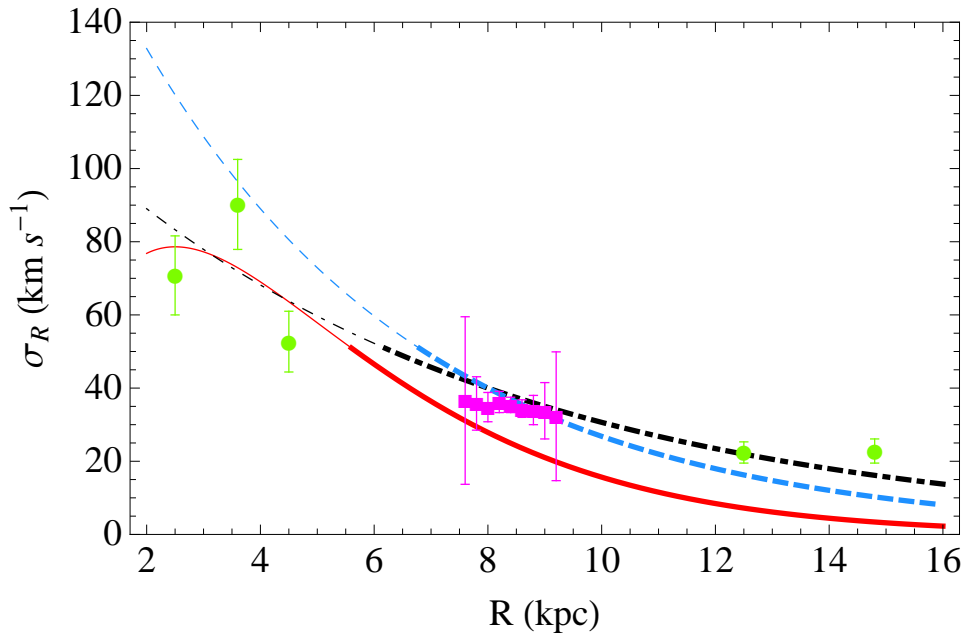}

\caption{\rw{The radial velocity dispersion profiles of the three models outlined in
Tables~\ref{tbl:Values} and \ref{tbl:Models}. The curves represent $\sigma_R\propto
\Sigma(R)^{1/2} \propto e^{-R/2R_d}$ (\kch{Model~$\Sigma$ - }dashed,
blue), $\sigma_R\propto e^{-R/3 R_d}$ (\kch{Model~W -} dot-dashed,
black), and $\sigma_R(R)\propto R\,e^{-R/R_d}$ normalised so that
$Q=1.5$ \kch{(Model~Q - thick, red) for} a flat rotation curve.  For
reference, we have plotted data  for the Milky Way disc for K-giant stars (green, circles) \citep{LF89} and for 
thin disc stars from the} 
RAVE survey (selected based on kinematics and location from the plane  $-0.5<z \rm{(kpc)}\leq -0.3$; magenta, squares\kd{)  \citep{Pasetto12}.}
Thinner line types  indicate a radial velocity dispersion $\sigma_R > 50$~km~s$^{-1}$\kch{, where the epicyclic approximation may break down and we have not tested our \kd{capture} criterion}.}

\label{fig:RMEII_sigR_exp_vs_Q}
\end{center}
\end{figure}

The radial profiles for $\sigma_R(R)$ for each \rw{of these three
prescriptions} are illustrated in
Figure~\ref{fig:RMEII_sigR_exp_vs_Q}.  The curves show Model~$\Sigma$
(dashed, blue), \rw{Model~Q (thick, red), and Model~W (dot-dashed,
black), normalised as in Table~\ref{tbl:Values}.  We also show
estimated values for the Milky Way disc from observations of K-giant
stars (green, circles; \cite{LF89} and thin \kch{disc} stars
from the RAVE survey (magenta, squares) \citep{Pasetto12}.  The
$Q=1.5$ \rw{curve lies below the data} for much of the disc, implying
that the adopted normalisation is not well-matched to the Milky
Way. This does not affect the \it{trends} that are the main focus of
our investigation.}


\rw{The capture criterion (equation~\ref{eqn:CaptureCriterion}) was derived using 
the epicyclic approximation} and thus implicitly assumes small
amplitude random motions.  \rw{We established the validity of this 
criterion  for any single component of stellar random velocity
up to $\sim 50$~km~s$^{-1}$ \citep{DW15} but we did not test it
above this \kch{value}.  We therefore use thick/thin lines to distinguish these two regimes in Figure~\ref{fig:RMEII_sigR_exp_vs_Q} and subsequent figures.} 

\subsection{Orbital Angular Momentum Distributions}\label{sec:AMDistribution}

The distribution of orbital angular momentum in the {disc} plays a critical role in determining \rw{which stars can be in trapped orbits, and in this section we evaluate this distribution for each of our models\footnote{The distribution of orbital angular momentum in the \kch{disc} may be \rw{rearranged at some later time.  However,} \cite{SB02} and \cite{Sellwood14} \kch{argue} that the net change in the orbital angular momentum distribution \kch{due to radial \kch{migration}} is nearly zero for any distribution function that has a shallow gradient in orbital angular momentum, $\partial f/\partial L_z$, near corotation.}}.

\rw{We assume for models $\Sigma$ and W that a population characterised by radial velocity dispersion $\sigma_R$ at radius $R$ will have a mean azimuthal velocity $\langle v_\phi \rangle$ that lags the circular orbital velocity by $v_a$ (equation~\ref{eqn:AD}) and mean guiding centre radius interior to R. }

We calculate the \kch{radial} distribution of orbital angular momentum by dividing the disc into annuli of width $\delta R$ and evaluating  the angular momentum content in each annulus as follows: 
\begin{equation}\label{eqn:LzIntegral}
L_{z}(R_i)=\int_{-\infty}^{\infty} \int_{-\infty}^{\infty} \int_0^{2\pi}  \int_{R_i}^{R_i+\rw{\delta R}}
\rw{\kd{R \, v_\phi}} \, f(\mathbf{x},\mathbf{v})\, R\, dR\,d\phi\,dv_R\,dv_\phi,
\end{equation}
where \rw{we have assumed that all stars have unit mass (for an \kd{individual} star, $L_{z,*} = R_*\,v_{\phi,*}$).}
The orbital angular momentum in each annulus of an exponential \kch{disc} with a flat rotation curve that is entirely composed of stars in circular orbits is given by,
\begin{equation}\label{eqn:LzcIntegral}
L_{z,c}(R_i)=2\pi\, v_c  \int_{R_i}^{R_i+\rw{\delta R}}\, \Sigma(R)\, R^2\, dR.
\end{equation}

\begin{figure}
\begin{center}
\includegraphics[scale=1]{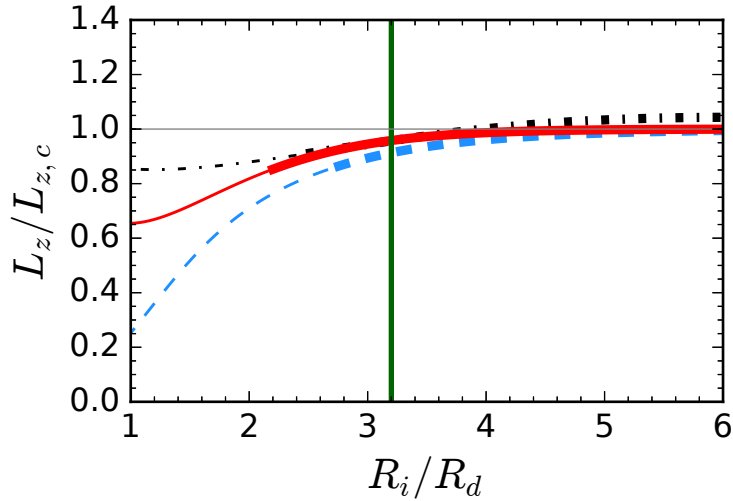}

\caption{Radial distribution of orbital angular momentum for the
following three \kch{disc} models: Model~$\Sigma$ (blue, dashed),
Model~W (black, dot-dashed), and Model~Q (red, solid), where
$R_d=2.5$~kpc.  \rw{The angular momentum content of each annulus} is
normalised \kch{by the angular momentum in the same annulus for} an
exponential \kch{disc} that is entirely composed of stars in circular
orbits ($L_z(R_i)/L_{z,c}(R_i)$).  Each annulus has \rw{width 
$\delta R=0.1$~kpc (sufficiently narrow that the curves appear continuous)}.  A vertical, green line indicates the radius of
normalisation ($R_0=8$~kpc) \kch{for the appropriate surface density
and radial velocity dispersion profiles described in \S\ref{sec:DF}.}
Thin lines indicate \rw{where} each model has radial
velocity dispersion $\sigma_R > 50$~km~s$^{-1}$.}
\label{fig:RMEII_LzR}
\end{center}
\end{figure}

\rw{The radial distribution of orbital angular momentum, $L_z(R_i)$, for
each model of Table~\ref{tbl:Models} is shown in Figure~\ref{fig:RMEII_LzR}, normalised in each annulus by the corresponding purely circular orbit model ($L_{z,c}(R_i)$).  In adopting this  normalisation 
we are implicitly assuming that the 
integral of each distribution function over velocity space gives
a surface density profile that is well approximated by an exponential
(equation~\ref{eqn:SurfaceDensityProfile}).  The  line types  in Figure~\ref{fig:RMEII_LzR} correspond to those in 
Figure~\ref{fig:RMEII_sigR_exp_vs_Q}, and again thin} lines
represent regions where the radial velocity dispersion
$\sigma_R>50$~km~s$^{-1}$ \rw{and underlying assumptions may be expected to break down}.

The radial profile for the mean orbital angular momentum goes as
$\langle L_z (R) \rangle \propto R\, \Sigma(R) \,\langle v_\phi
(R)\rangle$, where $\langle v_\phi (R)\rangle$ is the mean azimuthal
velocity of the \rw{stars at radial coordinate $R$.}
The radial Jeans equations \kch{show}
that for a given \rw{scalelength,} populations that have higher random
\kch{orbital energy have lower} azimuthal streaming velocity (see
\S\ref{sec:CapturedFractionSigNotRadDep} for a brief discussion).  It
follows that the normalised orbital angular momentum distribution for
each model decreases toward the \rw{inner regions as the} velocity
dispersion increases.

\kch{The differences between the shapes of the curves for each orbital angular momentum profile in Figure~\ref{fig:RMEII_LzR} can be understood \rw{in trems of the} differences in each model's radial velocity dispersion profile and radial surface density profile.
For the two models (Model~$\Sigma$ and Model~Q) that adopt a Gaussian velocity distribution ($P_G(\mathbf{v})_\mathbf{x}$), the mean angular momentum profile is approximated by $\langle L_z (R) \rangle \sim R\, \Sigma(R) \,(v_c-v_a(R))$, where $v_a\propto \sigma_R^2$.
By construction, these two models also have} an exponential surface density profile (equation~\ref{eqn:SurfaceDensityProfile})\kch{.}  Both of these models also have a radial velocity dispersion profile ($\sigma_R(R)$) that goes to zero at large radii, and thus each normalised orbital angular momentum profile converges to unity with increasing radius.   
We also assume an exponential form for the \kch{normalisation of the surface density profile for Model~W }though this is not strictly self-consistent.  An analysis of the robustness of approximating the integral of $f_D$ over velocity space, $\Sigma_D(R)$, as an exponential with scale length $R_d$ is given by \cite{Dehnen99b}.  
We note that at radii $R_i/R_d\gtrsim 4$, the surface density profile for $\Sigma_D(R)$ has values greater than \rw{those of} the exponential surface density profile used in \rw{the} normalisation and \rw{those of} models that adopt a Gaussian velocity distribution; \rw{this is consistent with the fact that} the normalised orbital angular momentum profile for Model~W has values greater than unity at $R_i/R_d\gtrsim 4$.

\section{The Fraction of Stars in Trapped Orbits}\label{sec:CapturedFraction}

In order to investigate how stellar kinematics affect the fraction of
a population that is in trapped orbits \rw{(hereafter denoted by
$\mathcal{F}$),} we apply the capture criterion
(equation~\ref{eqn:CaptureCriterion}) to our models of \kch{disc}
populations.  In the method described below, we solve numerically for
the fraction of stars in trapped orbits given a particular \kch{disc}
potential and form for the distribution function. We adopt the
unperturbed, axisymmetric stellar distribution function, then impose a
given spiral perturbation and evaluate for given value of the
\rwqq{phase space coordinates} whether or not the capture criterion
(equation~\ref{eqn:CaptureCriterion}) is met. \rwqq{As we will see, stars
with phase space coordinates not coincident with the corotation
resonance can still be trapped, and the use of the unperturbed
axisymmetric distribution in the calculation of the trapped
fraction needs to be checked. \citet{Bovy15} investigated the response of both cold and
warm 2D discs to a non-axisymmetric perturbation and found that the
warmer the disc, the less it responded. \kjd{Azimuthal (their tangential)} velocity is the most important phase space
coordinate in our analysis  (our proxy
for orbital angular momentum) and this is less affected than is the
radial velocity.  Bovy found that for typical weak non-axisymmetric
perturbations, such as an elliptical disc or bar (the latter grown either abiabatically or rapidly), the mean tangential
velocity at a given location is perturbed by less than 5\%, in a disc
with \kjd{initial ratio} $\sigma_R/v_c < 0.2$ (see his Figures 26, 29 and
30). The amplitude of the response was seen  to increase
approximately linearly with the amplitude of the imposed potential. We  expect that while our
results will systematically over-predict the
fraction of stars in trapped orbits, the trends we find with velocity dispersion and location will be valid.}

The first step in the evaluation \rwqq{of the trapped fraction} is to find the fraction of stars
within each annulus, $R_i+\delta R$, that satisfies the capture
criterion. The numerical integration for a given annulus is carried
out assuming that the fraction is \kdd{constant across the width of} the annulus, and only
one value of the radial coordinate is used, \kdd{which corresponds} to the
inner radius of the annulus\footnote{We validated this assumption
provided the width is less than a typical epicyclic excursion.}.
 
We integrate the distribution function at $R$ over the remaining phase-space -- azimuthal coordinates, $\phi$, and velocity space coordinates, $v_R$ and $v_\phi$, -- with limits of integration set by the region of phase-space that satisfies the capture criterion.  We set absolute limits of integration in velocity space $\langle v_i \rangle\pm 100$~km~s$^{-1}$, where $i$ represents the radial or azimuthal components of the velocity vector.
The solution to this integral, divided by the integral of the distribution function over all phase-space at $R$, is the fraction of stars in trapped orbits within a given annulus. 
We calculate the fraction of stars in trapped orbits \kch{for all annuli within $R=R_{CR}\pm4$~kpc for} each choice of initial conditions.  We thus find the radial distribution for the fraction of stars in trapped orbits per annulus, \kch{$\mathcal{F}_{R}\equiv \mathcal{F}(R)$,} where the subscript \lq\lq $R$" indicates evaluation at the radial position $R$ associated with that annulus.

We perform all integrations using the \textit{Wolfram Mathematica v.8.0.4.0} software function \texttt{NIntegrate} with the parameters \texttt{AccuracyGoal} set to 8 significant digits and \texttt{MinRecurrsion} set to 8.  We chose the values of these parameters to ensure convergence to a solution within a reasonable amount of time.\footnote{This is on order of a \kch{half hour} of computing time per set of initial conditions with a \kch{4~GHz} processor.}  
\kch{We set the bounds of integration to be the region of phase-space that satisfies the capture criterion using the \texttt{Boole} function.}

Our primary focus here is to understand trends in the the relationship between the fraction of stars initially in trapped orbits and the stellar velocity dispersion ($\sigma$) for the population \rw{and so the \it{relative} trapped fraction between different models is of most interest}.  In the process we explore how the fraction of stars in trapped orbits within a chosen radial range, 
\begin{equation}\label{eqn:FDeltaR}
\mathcal{F}_{\Delta R}\equiv \int_{\Delta R} \mathcal{F}(R) \,dR ,
\end{equation}
depends on the \kd{adopted} functional form of the distribution function and the parameter values of the spiral potential.\kd{\footnote{\kd{Equation~\ref{eqn:FDeltaR} implicitly includes an integration of the chosen distribution function over all phase-space, excepting radial position, as described at the beginning of \S\ref{sec:CapturedFraction}.}}}
The subscript \lq\lq $\Delta R$" indicates that the fraction of stars in trapped orbits is evaluated by integrating over a specific radial range, $\Delta R$\kd{, of annuli, where the value for $\mathcal{F}_i$ within each annulus has been evaluated over the non-axisymmetric azimuthal distribution of stars in trapped orbits (as described above).}  \kch{We henceforth refer to the fraction of stars in trapped orbits within a chosen radial range ($\mathcal{F}_{\Delta R}$) as the \lq\lq integrated fraction\rq\rq.}
\kdd{We do not evaluate the time evolution of
the integrated fraction, and therefore refer to the fraction of stars that meets
the capture criterion for a given model as the \lq\lq initial"
fraction.   In \S\ref{sec:UpperLimit}, we argue that the initial fraction over-predicts the fraction of stars that could migrate radially.}

\kd{For the entirety of this section we adopt either a radially fixed velocity dispersion profile and Gaussian distribution function ($f_G(\mathbf{x},\mathbf{v})$) or our simplest fiducial model, Model~$\Sigma$.  In Section~\ref{sec:CapFracInModels}, we use insights gained from these explorations to \kdd{investigate} the full fiducial models, Models~$\Sigma$, Q, and~W, of the galactic disc.}


\subsection{\kch{Integrated} Fraction for a \rw{Fixed} Radial Velocity Dispersion \rw{and Gaussian Distribution Function}}\label{sec:CapturedFractionSigNotRadDep}

The fraction of \rw{stars in a given disc population} that meets the capture criterion
at a given \rw{spatial} coordinate \rw{depends on the population's velocity dispersion}\footnote{This statement should not be confused
with the fact that whether or not an \textit{individual} \kch{disc}
star is in a trapped orbit is largely independent of its random
orbital energy and determined primarily by its orbital angular momentum (or circular frequency)\citep{DW15}.}. In this subsection \kch{we use \rw{a 
simple model to} gain physical insight into the \rw{form of this dependence.}
We \rw{adopt the Gaussian} phase space distribution function
$f_G(\mathbf{x},\mathbf{v})$ (equation~\ref{eqn:fGfull}, described in
\S\ref{sec:fG}), a radial velocity dispersion ($\sigma_R$) that is
constant \rw{within the disc} and a fixed ratio \kch{between the} radial and
azimuthal velocity dispersions, $\sigma_\phi/\sigma_R$.  The disc and spiral pattern are assumed to have the parameter values given in Table~\ref{tbl:Values}, with $R_{CR}=8$~kpc.}

\subsubsection{Radial distribution of the fraction of stars in trapped orbits}\label{sec:RadialDistributionSig}

\begin{figure}
\begin{center}
\includegraphics[scale=1]{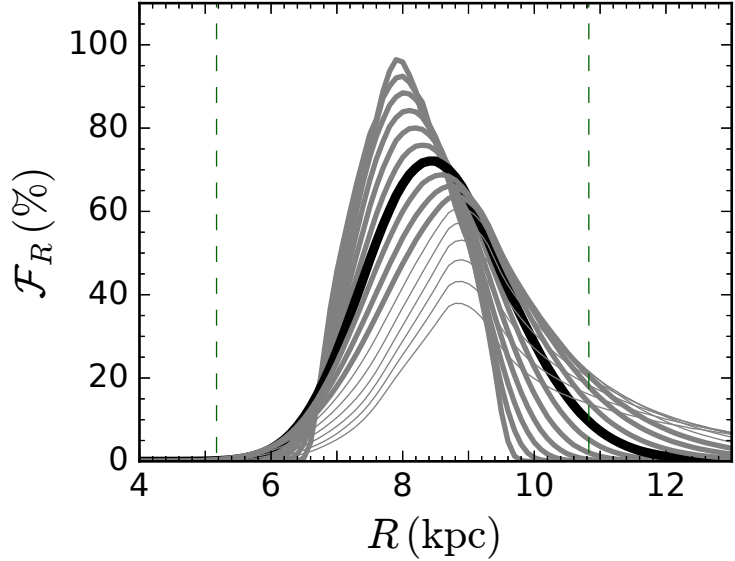}
\caption{Radial distributions for the fraction of stars in trapped orbits, evaluated using the method described in the introduction of \S\ref{sec:CapturedFraction} and assuming a radially \rw{invariant} radial velocity dispersion profile.  The \rw{Gaussian form of the velocity} distribution function ($f_G$) has \rw{been adopted} and the perturbing spiral potential has $R_{CR}=8$~kpc and parameter values given by Table~\ref{tbl:Values}.  \rw{The assumed value of the radial velocity dispersion ranges from 5~km~s$^{-1}$ to} 80~km~s$^{-1}$ in $5$~km~s$^{-1}$ increments.  For reference, the \rw{curve} for the \kch{model that} assumes $\sigma_R=35$~km~s$^{-1}$ is in bold.  Thin lines indicate \rw{results for}  radial velocity dispersion $\sigma_R > 50$~km~s$^{-1}$.  \kref{For reference, the thin, vertical, long-dashed (dark green) lines mark the inner and outer Lindblad resonances.}}
\label{fig:RMEII_CaptureDistribution_ADvsnoAD}
\end{center}
\end{figure}

Figure~\ref{fig:RMEII_CaptureDistribution_ADvsnoAD} shows the fraction
of stars in trapped orbits at a range of radial coordinates \kch{for a
given} radial velocity dispersion, $\sigma_R$, ranging between $5-80$~km~s$^{-1}$, in
$5$~km~s$^{-1}$ increments.   We devote the rest of this
subsection to \rw{identifying} the physics that determines the shapes of
these \rw{curves.  The intuition developed will be useful when we investigate 
more complex models later in this paper.}

In the limiting case that $\sigma_R \rightarrow 0$ \rw{and all} orbits
approach circular (in an axisymmetric potential), \rw{the radial distribution of
the fraction of stars in trapped orbits} saturates at
$100\%$ \kch{within the capture region and goes to zero at coordinates
outside the capture region.  An approximation for the integrated
fraction ($\mathcal{F}_{\Delta R}$) \rw{in this limit can be obtained from the ratio of the area of the  capture region, $A_{cap}$, to 
the area of the annulus, $A_{ann}$, that spans the maximum radial range
of the capture region. For low-amplitude spiral arms ($\epsilon_\Sigma
\lesssim 0.3$) that have a low pitch angle
($\theta\lesssim 25^\circ$) the capture region follows closely the closed contours around the maxima in
the effective potential (as seen 
in~Figure~\ref{fig:RMEII_CaptureRegion}(b)) and these in turn are well approximated by a
sinusoidal function in the azimuthal direction (see
equation~\ref{eqn:SpiralPotential}).  Under
these conditions the area of the
capture region may be evaluated as:}
\begin{equation}
A_{cap} = \int_0^{2\pi} \int_{R_{CR} -(R_{CR}-R_{min})|\cos\phi |}^{R_{CR} +(R_{max}-R_{CR})|\cos\phi |} \,R\,dR\,d\phi,
\end{equation}
\rw{where $R_{min}$ and $R_{max}$ denote the minimum and maximum
radial coordinates of the capture region.  Making the reasonable
assumption that the capture region is symmetric around  corotation, and
being careful to note that the limits of integration involve $|\cos
\phi|$, it is straightforward to} show that 
$A_{cap}=4 R_{CR}(R_{max}-R_{min})$.  The area of the
annulus enclosing the capture region \rw{($A_{ann}$) is} $A_{ann} =
2\pi R_{CR}(R_{max}-R_{min})$. The integrated fraction of stars in the
annulus containing the capture region is therefore
$\mathcal{F}_{\Delta R}=2/\pi \approx 64\%$.  \rw{Values higher than 
 $2/\pi$ can be obtained if the annulus over which the trapped fraction is integrated is narrower than the capture region, with the integration fraction approaching 100\% around corotation, where the capture region spans all azimuthal
coordinates. These limiting cases will be used} to inform our interpretation of results that
appear later in this paper.}


\rw{The rms  maximum radial excursion from the guiding centre radius, due to epicyclic motion, for a population with radial velocity dispersion $\sigma_R$ is given  by}
\begin{equation}\label{eqn:X}
X(R) = \dfrac{\sigma_R(R)}{\sqrt{2}\,\kappa(R)}
\end{equation}

\rw{Thus in a disc with constant {radial} velocity dispersion and a flat rotation curve, the rms maximum radial epicyclic excursions scale as $X \propto R$. This typical maximum epicyclic excursion is a measure of how far from the limit of the \kd{capture} region a star can be physically and still have its guiding centre within the capture region and be in a trapped orbit. } 

\rw{The curve in Figure~\ref{fig:RMEII_CaptureDistribution_ADvsnoAD}
for the lowest velocity dispersion ($\sigma_R = 5$~km~s$^{-1}$) peaks
close to the corotation radius (8~kpc), and} 
the fraction of stars in trapped orbits in annuli away from corotation
approximately equals the fraction of azimuthal coordinates that are
within the capture region.   \rw{Populations of stars with
higher values of their velocity dispersion, on  more
eccentric orbits, have a greater range of radial positions outside the capture
region and therefore} the distribution of stars in trapped orbits spreads over a greater
radial range \rw{as seen in  Figure~\ref{fig:RMEII_CaptureDistribution_ADvsnoAD}}.
\kref{High velocity dispersion populations can have a distribution of stars in trapped orbits that extends beyond the inner and outer Lindblad resonances (radii at which $\kappa=\pm m(\Omega_p-\Omega_c)$ -- marked with long-dashed in Figure~\ref{fig:RMEII_CaptureDistribution_ADvsnoAD}) where stars are likely to be scattered out of trapped orbits.}

\begin{figure}
\begin{center}
\includegraphics[scale=1.2]{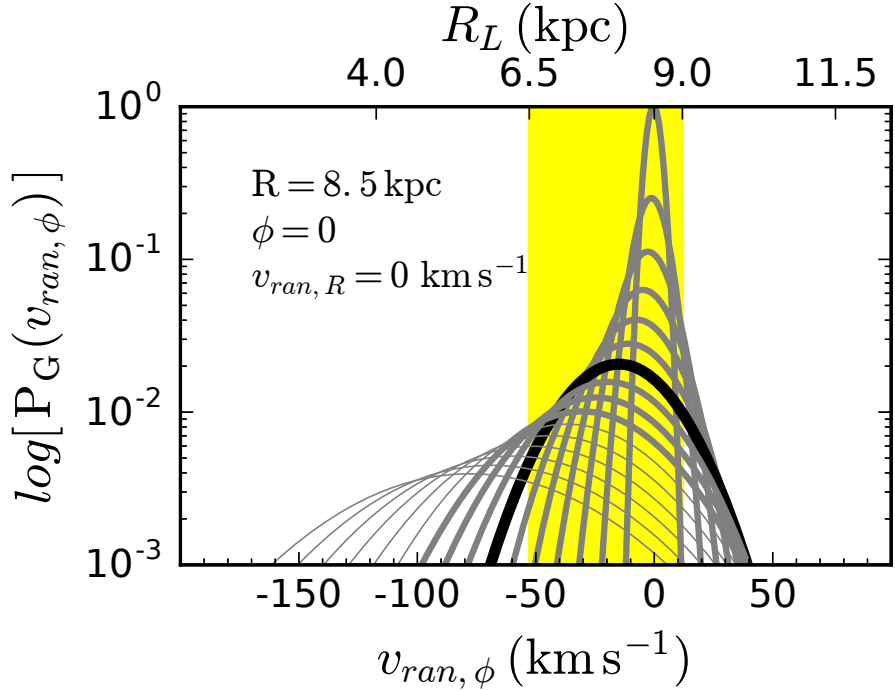}
\caption{\rw{Distributions of the random azimuthal velocity
($v_{ran,\phi}$) for stars observed at $R=8.5$~kpc, and $\phi=0$
(between the spiral arms, just beyond corotation at $R_{CR}=8$~kpc)
with $v_{ran,R}=0$~km~s$^{-1}$. The underlying model is as
Figure~\ref{fig:RMEII_CaptureDistribution_ADvsnoAD}.}  The vertical
axis shows the probability $P_G(\mathbf{v})_\mathbf{x}$ on a
logarithmic scale.  Each curve represents \rw{a stellar population
with a radial velocity dispersion
($\sigma_R$) \rw{in the range} $5-80$~km~s$^{-1}$, in $5$~km~s$^{-1}$
increments.  The \rw{thick (black) curve is for } radial
velocity dispersion $\sigma_R=35$~km~s$^{-1}$.  The shaded (yellow)
area indicates the range of azimuthal velocities that meet the capture
criterion.  The upper horizontal axis shows the guiding center radius
($R_L$) associated with $v_{ran,\phi}$.  Due to asymmetric drift, the
peak in the random azimuthal velocity distribution shifts toward
slower rotation with increasing velocity dispersion.  In \rw{a corresponding} manner, the peak of the probability distribution of guiding centre
radii shifts toward radii that are closer to the galactic centre than
the coordinate radius, $R$. Thin lines indicate model realisations
that use radial velocity dispersion $\sigma_R > 50$~km~s$^{-1}$.}}
\label{fig:RMEII_VelocityDistribution}
\end{center}
\end{figure}

\rw{The location of the peak in the trapped fraction also depends on
velocity dispersion, shifting to larger galactocentric radii for
kinematically hotter populations. This reflects the shift in mean
azimuthal velocity, represented by our simple asymmetric drift formula
(equation~\ref{eqn:AD}\kd{)}.  To understand this, first consider how the
contribution to the \kch{integrated fraction} from a stellar
population at given coordinates $\mathbf{x}$ depends on velocity
dispersion. This is illustrated in
Figure~\ref{fig:RMEII_VelocityDistribution}, which shows the
distribution of random velocities in the azimuthal direction for stars
observed at coordinates $R=8.5$~kpc, $\phi=0$ (just outside
corotation, between the spiral arms), and with $v_{ran,R}=0$. The different curves are for radial velocity dispersions in the range $5-80$~km~s$^{-1}$, in
$5$~km~s$^{-1}$ increments. 
The peak of the distribution of $v_{ran,\phi}$ and of 
the corresponding distribution of guiding centre radii, $R_L$, shift} to lower values with higher radial velocity
dispersion.  The yellow (shaded) region \rw{indicates those} values of $v_{ran,\phi}$
(and $R_L$) that meet the capture criterion.  

\begin{figure}
\begin{center}
\includegraphics[scale=1]{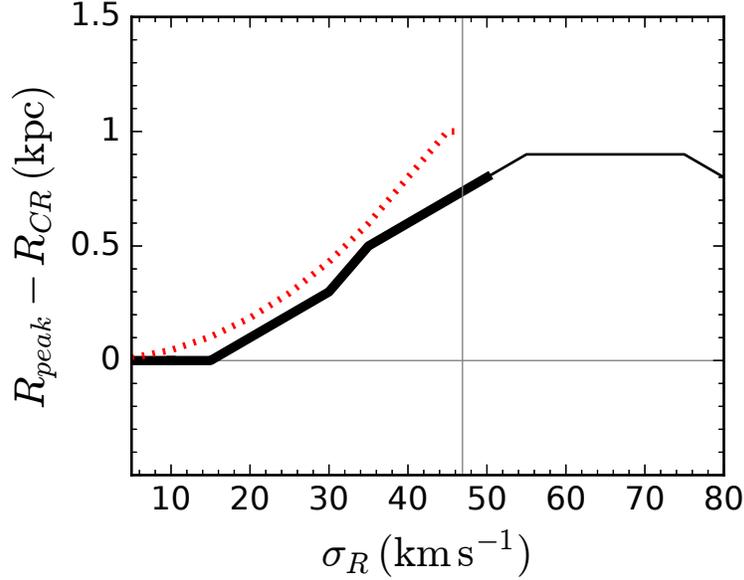}
\caption{Radial offset of the \rw{location of the} peak ($R_{peak}$)
in the distribution of the fraction of stars in trapped orbits
($\mathcal{F}_R$) from the radius of corotation ($R_{CR}$) as a
function of the assumed fixed radial velocity dispersion
($\sigma_R$). The \rw{solid} (black) curve shows $R_{peak}-R_{CR}$
\kch{calculated from the same models} shown in
Figure~\ref{fig:RMEII_CaptureDistribution_ADvsnoAD}, while the \kd{dotted}
(red) curve shows the values predicted by
equation~\ref{eqn:RPeakPredict} \rw{(with $v_c = 220$~km~s$^{-1}$ and $R_{CR} = $8~kpc), up to the critical velocity dispersion $\sigma_{crit}$ from 
equation~\ref{eqn:SigmaCrit} beyond which this relation is invalid (indicated by the thin vertical line).}
\rw{ Again, results for values of the radial velocity dispersion greater than 50~km~s$^{-1}$ are shown with thin lines}.  The increase in the shift of $R_{peak}$ from
$R_{CR}$ \rw{as  asymmetric drift increases with increasing values of the velocity dispersion is apparent.}}
\label{fig:RMEII_Offset_AD}
\end{center}
\end{figure}

\rw{The radial coordinate of the peak in the radial distribution of the fraction of stars in
trapped orbits ($R_{peak}$ \kd{-- in Figure~\ref{fig:RMEII_CaptureDistribution_ADvsnoAD}}) corresponds to the radius} at which
the mean guiding centre radius for the stellar population \rw{of interest} equals the
radius of corotation.
\kd{The radius of corotation is the only radius at which all azimuthal coordinates are within the capture region, modulo a slight radial broadening for very strong spiral patterns (compare panels (b)~and~(c) in Figure~\ref{fig:RMEII_CaptureRegion}).  }
\rw{The mean guiding centre radius ($\langle R_L \rangle$) of stars
observed at coordinate $R$ reflects the mean angular momentum of the
stars, which in our prescription is set by the velocity dispersion,
through the asymmetric drift equation (equation~\ref{eqn:AD}). 
\kd{Thus, stellar populations with velocity dispersion approaching zero ($\sigma_R\rightarrow 0$), where the radial position of each star approaches its guiding centre radius ($R\rightarrow R_L$), have $R_{peak}\rightarrow R_{CR}$.}
$R_{peak}$ increases (moves to larger distances from the galactic
centre) with increasing velocity dispersion}.  We show \rw{the
dependence of $R_{peak}$ on the radial velocity dispersion in
Figure~\ref{fig:RMEII_Offset_AD}; the thick (black) curve represents
the offset $R_{peak}-R_{CR}$ for each value} of the radial
velocity dispersion shown in
Figure~\ref{fig:RMEII_CaptureDistribution_ADvsnoAD}.

\rw{These physical arguments suggest that the value of $R_{peak}$ can
be determined by finding the radius at which the mean guiding centre radius equals the
radius of corotation. We then have:}
$$\langle R_L \rangle=R \langle
v_\phi\rangle/v_c$$
so that 
$$R_{peak}\,\langle v_\phi\rangle=R_{CR}\,v_c$$ 
and 
\begin{equation}\label{eqn:RPeakPredict}
R_{peak}-R_{CR} = R_{CR} \left(\dfrac{v_c}{v_c-v_a}-1 \right).
\end{equation}



\rw{The predicted relation from Equation~\ref{eqn:RPeakPredict} is
  shown in Figure~\ref{fig:RMEII_Offset_AD} (dashed, red curve)
  together with the results from our models (solid, black curves). The
  relation from Equation~\ref{eqn:RPeakPredict} is a good match to the
  models,} up to some critical value for the velocity \rw{dispersion,
  above} which the offset $R_{peak}-R_{CR}$ \rw{reaches a
  plateau. This maximum value for the offset is obtained when the
  radial coordinate for the population of stars that has a mean
  guiding centre radius equal to the radius of corotation is greater
  than, or equal to,} the maximum radius of the capture region
  ($R_{max}$).  \rw{Beyond this limit, stars in the 
{\it peak}\/
  of the velocity distributions no longer meet the capture
  criterion and  the contribution to the trapped fraction comes from stars in the broad, low-amplitude wings.} Thus this \rw{critical value of the velocity dispersion  marks the} boundary
  above which the dominant contribution to the fraction of stars in
  trapped orbits changes from stars that have both radial positions
  and guiding centre radii inside the capture region to stars with
  guiding centre radii only inside the capture region.  However since
  the capture region does not have constant radial width \rw{(see Figure~\ref{fig:RMEII_CaptureRegion})}, the
  transition between these two regimes is not abrupt.

\rw{An expression for this critical value of} the radial velocity
dispersion, $\sigma_{crit}$, \rw{may be obtained by considering 
the  stellar population located at
the maximum radius of the capture region and requiring that the mean angular momentum (per unit mass) of this population, given by   $\langle L_z\rangle = R_{max\,}\langle
v_\phi\rangle$,  be equal to that of a star in a circular
orbit at corotation (i.e.~$\langle L_z\rangle =R_{CR}\,v_c$).  } Given this
\rw{requirement} and our prescription for asymmetric drift
(equation~\ref{eqn:AD}), \rw{the critical velocity dispersion may be expressed as:}
\begin{equation}\label{eqn:SigmaCrit}
\kch{\sigma_{crit} = \sqrt{\kch{C}\,v_c\left(1-\dfrac{R_{CR}}{R_{max}}\right)}}
\end{equation}
\rw{The value of $\sigma_{crit}$ for the parameter values of the model in  Figure~\ref{fig:RMEII_Offset_AD} is indicated by the  thin, vertical line in that figure.}

\subsubsection{\kch{Definition of} the radial range of evaluation}\label{sec:DefiningRadialRange}

The fraction of stars initially in trapped orbits \rw{varies with} 
radius and it is \rw{important  to quantify}  how the integrated
fraction \kch{($\mathcal{F}_{\Delta R}$) depends upon} the radial
range \kch{over which it is evaluated}.  We define \kch{this} radial
range\kch{, $\Delta R$,} \rw{with two different approaches, the first a fixed physical width and the second depending on the shape of the distribution of the trapped fraction.} 
Two of the measures \rw{of the trapped fraction, denoted $\mathcal{F}_1$ and $\mathcal{F}_2$}, \kch{are
evaluated using} \rw{annuli centred at the corotation radius and of fixed width  $\Delta R_{1} = 1$~kpc and 
$\Delta R_{2} = 2$~kpc, respectively}.  We also \rw{evaluate} \kch{three measures that
\kref{do not have a fixed centre, rather these radial ranges for evaluation}
are \rw{defined in terms of 
the radial} distribution of stars in trapped orbits.  For these,
\rw{denoted by $\mathcal{F}_{5\%}$, $\mathcal{F}_{25\%}$, and $\mathcal{F}_{FWHM}$, the requirement is that in every
annulus (each of width $\delta R$, given in Table~\ref{tbl:Values}) within the
radial range of evaluation, the fraction of stars in trapped orbits \kch{must be greater than}
$5\%$ (defining the radial range $\Delta R_{5\%}$), $25\%$ ($\Delta R_{25\%}$)\kch{,} and half
the maximum fraction captured ($\Delta R_{FWHM}$), respectively.}}

\begin{figure}
\begin{center}
\includegraphics[scale=1]{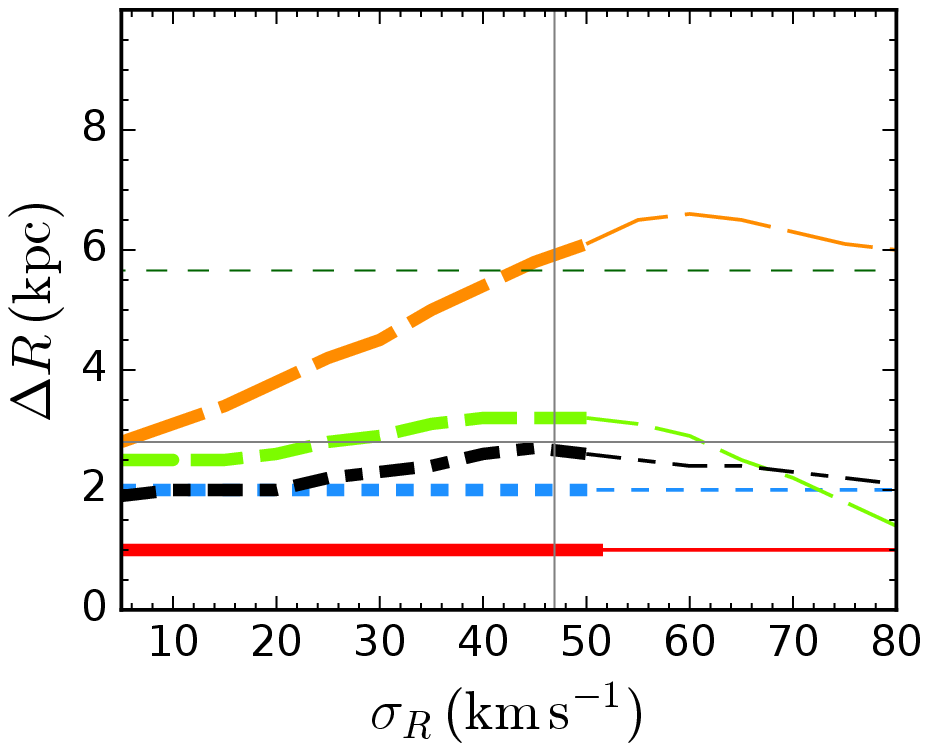}
\caption{The \rw{width of the} radial range ($\Delta R$) used to
evaluate the integrated fraction of stars ($\mathcal{F}_{\Delta R}$)
for $R_{CR}\pm0.5$~kpc ($\Delta R_{1}$ -- horizontal, solid, red),
$R_{CR}\pm1$~kpc ($\Delta R_{2}$ -- horizontal, short-dashed, blue),
and the radial range within which the annular fraction of stars in
trapped orbits is $>~5\%$ ($\Delta R_{5\%}$ -- long-dashed, orange),
$>~25\%$ ($\Delta R_{25\%}$ -- dashed, green), and greater than half
the maximum value ($\Delta R_{FWHM}$ -- dot-dashed, black).  \rw{Thinner segments of lines indicate}  radial velocity
dispersion $\sigma_R > 50$~km~s$^{-1}$.  
The measures of $\Delta R$ shown here are \rw{evaluated} for the same
model illustrated in
Figure~\ref{fig:RMEII_CaptureDistribution_ADvsnoAD}. For
reference, the thin, horizontal (dark-green) \rw{line indicates the
distance} between \kch{the inner and outer} Lindblad resonances
(dashed) \rw{and the horizontal grey line indicates the maximum width of the capture region}.   \rw{The} thin, vertical
line \rw{marks} the critical velocity dispersion, $\sigma_{crit}$, \rw{from equation~\ref{eqn:SigmaCrit}.} }
\label{fig:RMEII_DeltaR_AD}
\end{center}
\end{figure}

\rw{The widths of the above  radial ranges of interest are illustrated
in} Figure~\ref{fig:RMEII_DeltaR_AD}, for 
the model and \kch{parameter values of} 
Figure~\ref{fig:RMEII_CaptureDistribution_ADvsnoAD}. 
\rw{For low
values of the} velocity dispersion, \rw{$R_{peak}\rightarrow R_{CR}$ and the shape-dependent radial ranges approach an annulus centred on  corotation, as is prescribed for the
fixed measures $\Delta R_{1}$ and $\Delta R_{2}$. The variable widths generally are larger than the fixed widths, as seen in the figure.  Radial ranges narrower than the capture region will give higher integrated fractions than wider radial ranges and indeed can lead to higher values than the $\sim 64\%$ we estimated for widths equal to that of the capture region. \kref{Again, t}he widest radial ranges at high velocity dispersions can be comparable to the distance between the \kch{inner and outer} Lindblad resonances in
Figure~\ref{fig:RMEII_DeltaR_AD} and may include stars that are likely to be scattered out of trapped orbits.}

\begin{figure}
\begin{center}
\includegraphics[scale=1]{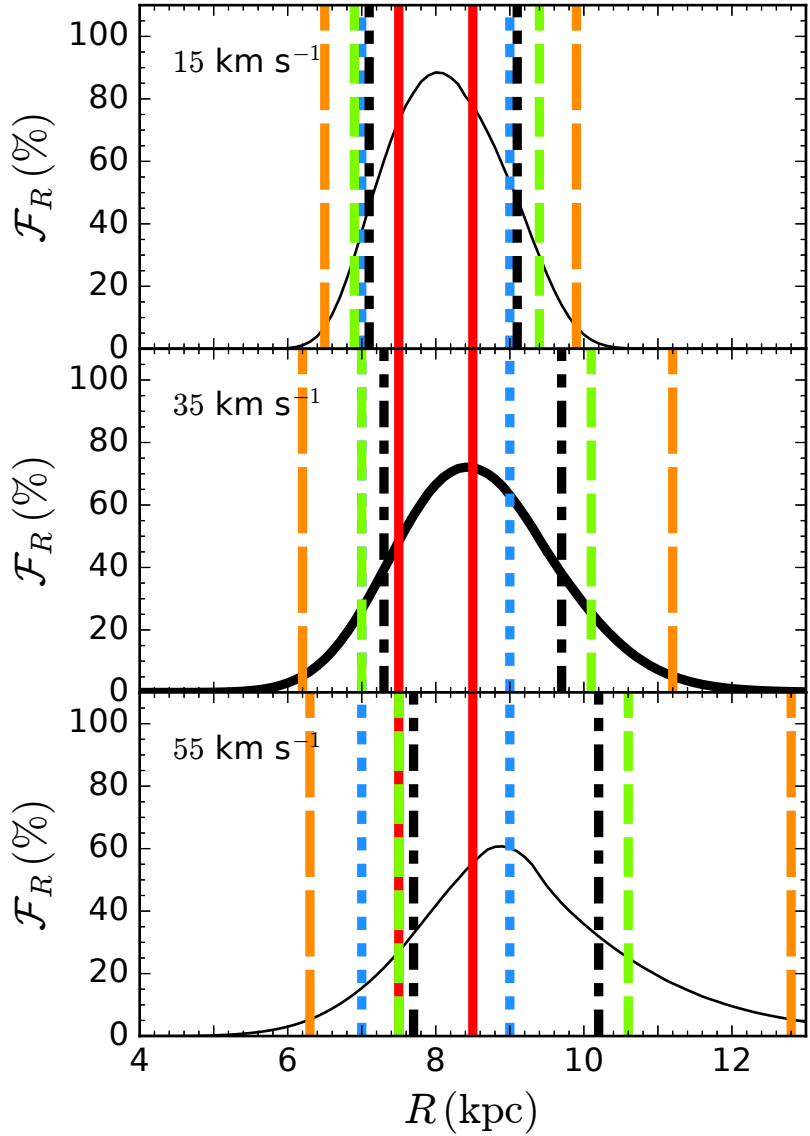}
\caption{\rw{Illustration of the various measures of the radial range for integration of the trapped fraction, applied to the distribution of trapped fraction \kref{with radial velocity dispersions $\sigma_R = \{15,35,55\}$~km~s$^{-1}$ as indicated in each panel}. The line types are taken from Figure~\ref{fig:RMEII_DeltaR_AD}.}}
\label{fig:RMEII_DeltaRR}
\end{center}
\end{figure}

\rw{An illustration of these various measures of the radial range of integration is given in Figure~\ref{fig:RMEII_DeltaRR}, applied to the curve of the distribution of the trapped  fraction for \kref{populations with $\sigma_R=\{15,35,55\}$~km~s$^{-1}$}, from  Figure~\ref{fig:RMEII_CaptureDistribution_ADvsnoAD}.}  \kref{For populations with high radial velocity dispersion (e.g.,~$\sigma_R=55$~km~s$^{-1}$ in Figure~\ref{fig:RMEII_DeltaRR}), the peak of the distribution of stars in trapped orbits may not be included in evaluations of the integrated fraction that use a fixed radial range ($\Delta R_{1,2}$). }

\subsubsection{\kch{Dependence of the integrated fraction} on velocity dispersion}\label{sec:IntegratedFractionSigRConstant}

\begin{figure}
\begin{center}
\includegraphics[scale=1]{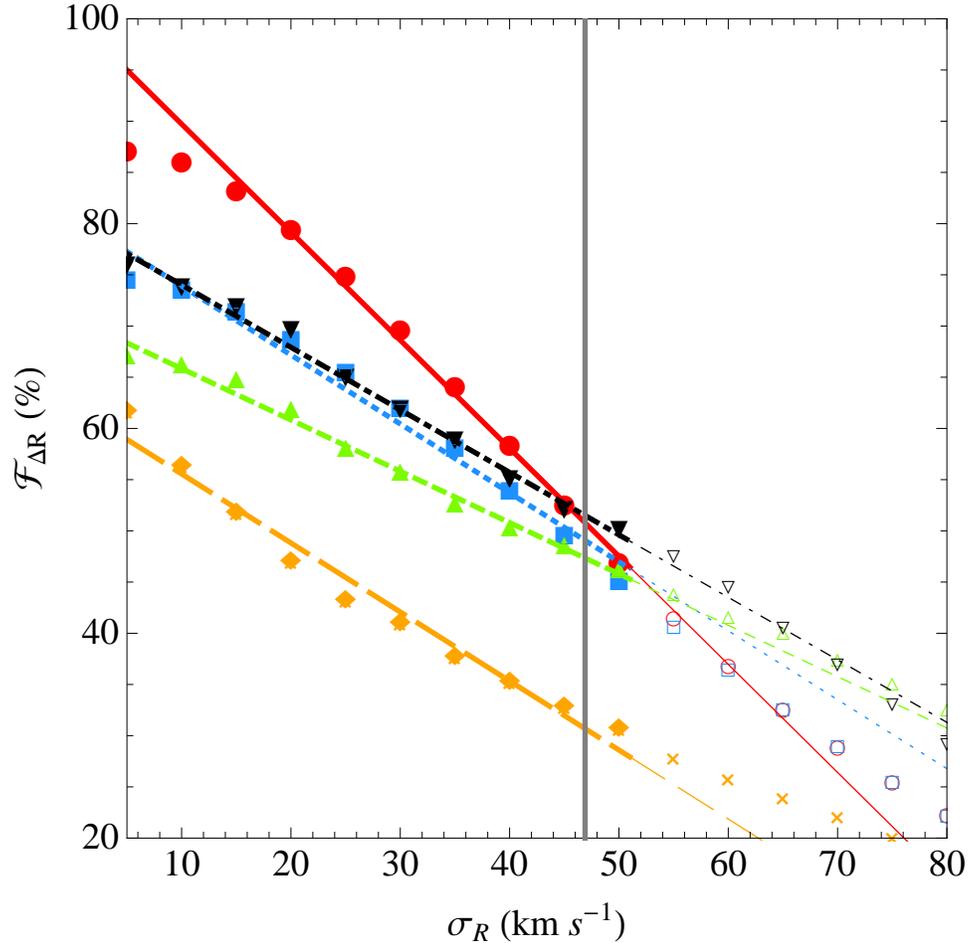}
\caption{The evaluations for
$\mathcal{F}_{1}$, $\mathcal{F}_{2}$, \kch{$\mathcal{F}_{5\%}$,
$\mathcal{F}_{25\%}$,} and $\mathcal{F}_{FWHM}$, 
marked with \kch{circles (red), squares (blue), diamonds (orange),
triangles (green), and downward-facing triangles (black)},
respectively. Best \kch{linear fits} for $\mathcal{F}_{1}$ (solid, red),
$\mathcal{F}_{2}$ (short-dashed, blue)\kch{, $\mathcal{F}_{5\%}$
(long-dashed, orange), $\mathcal{F}_{25\%}$ (dashed, green),} and
$\mathcal{F}_{FWHM}$ (dot-dashed, black).   The fitting parameters for
equation~\ref{eqn:CapturedFractionLinearFit} are given in
Table~\ref{tbl:FitLine}. The \rw{open symbols and thin lines  indicate an assumed} radial velocity
dispersion $\sigma_R > 50$~km~s$^{-1}$ and the vertical grey line indicates the critical value of the velocity dispersion from equation~\ref{eqn:SigmaCrit}. }
\label{fig:RMEII_FitSigma_AD}
\end{center}
\end{figure}

\rw{Figure~\ref{fig:RMEII_FitSigma_AD} shows the integrated
fractions $\mathcal{F}_1$, $\mathcal{F}_2$, $\mathcal{F}_{5\%}$,
$\mathcal{F}_{25\%}$, and $\mathcal{F}_{FWHM}$, together with the best-fit linear fits to the trends as a function of velocity dispersion, for the  model parameters of}  this subsection.  \rw{The open symbols and thin lines denote values of the radial velocity dispersion above 50~km~s$^{-1}$, and the 
thin,
vertical line indicates the critical velocity dispersion,
$\sigma_{crit}$, above which one might expect a change in behaviour.}


\rw{As seen in Figure~\ref{fig:RMEII_DeltaR_AD}, for velocity
dispersions below around 30~km~s$^{-1}$, the radial ranges used to
calculate the trapped fractions are narrower than the capture region,
with the exception of $R_{5\%}$. This explains why the various measures
of the trapped fraction are higher than the 64\% that was derived
earlier for the radial extent equal to the capture region. The values
of the measure $\mathcal{F}_{5\%}$ follows our expectations. This
measure includes a larger fraction of the disc and should be more
representative, modulo the caution that at large values of the
velocity dispersion the radial range becomes so wide that interactions
with the Lindblad resonances cannot be ignored. Indeed it is for this reason that we do not make an evaluation of the trapped fraction by integrating over the entire \kd{disc}, but rather limit the evaluation to the maximum radial range provided by $\Delta R_{5\%}$ and caution that $\mathcal{F}_{5\%}$ may include a significant population that is scattered out of trapped orbits, particularly at high velocity dispersions.
}

\rw{All measures of the trapped fraction show  a clear, approximately linear, decrease with increasing  velocity dispersion of the stellar population.}
\rw{The linear\footnote{We \rw{also explored} exponential,
polynomial, and Gaussian functional \rw{forms, but there is no  compelling physical reason to include these more complex relationships}.} fits shown in the figure used the function \texttt{LinearModelFit}} from \textit{Wolfram
Mathematica}'s built-in library\footnote{We set the fit \lq\lq method" to \lq\lq Automatic",
thus allowing \textit{Mathematica} to select from its library to
minimize fitting residuals.}  \rw{The values of the slope and zeropoint of the following linear fit are given in Table~\ref{tbl:FitLine}:}

\begin{equation}\label{eqn:CapturedFractionLinearFit}
\mathcal{F}_{\Delta R}(\sigma_R) = \xi_\mu \sigma_R +\xi_\beta ,
\end{equation}

We expect the values of the fitting constants, $\xi_{\mu}$ and
$\xi_{\beta}$, to depend \rw{on the underlying parameters such as the radius of 
corotation and the amplitude of the spiral potential, and explore this in section \S\ref{sec:FncDepOfTrapFrac}.}

\begin{table}
\begin{center}
\caption{\rw{The slopes and zero-points for the best-fit linear dependences of  the integrated fractions of stars in trapped orbits as a function of  radial velocity dispersions ($5\leq \sigma_R\leq 50$~km~s$^{-1}$). The underlying model} parameter values are given in Table~\ref{tbl:Values}.}
\begin{tabular}{ccc}
\hline
Integrated Fraction & Slope & $y$-intercept \\
$\mathcal{F}_{\Delta R}$ & $\xi_\mu$ & $\xi_\beta$ \\
\hline
$\mathcal{F}_{1}$        & -1.1 & 100 \\
$\mathcal{F}_{2}$        & -0.7 & 81 \\
$\mathcal{F}_{5\%}$    & -0.7 & 62 \\
$\mathcal{F}_{25\%}$  & -0.5 & 71 \\
$\mathcal{F}_{FWHM}$ & -0.6 & 80 \\
\hline
\end{tabular}
\label{tbl:FitLine}
\end{center}
\end{table}

\subsection{\kch{Integrated} \kch{F}raction for \kch{R}adially \kch{D}ependent \kch{V}elocity \kch{D}ispersion} \label{sec:CapturedFractionSigRadDep}

\rw{Our assumption in the last section that the radial velocity dispersion is constant with radius is reasonable provided that only a relatively narrow range of radius is being considered. }  We now explore \rw{how} the trends in the
integrated fraction shown in
\kch{\S\ref{sec:CapturedFractionSigNotRadDep}} are \rw{modified when a}
 radially dependent velocity dispersion profile
($\sigma_R=\sigma_R(R)$) \rw{is assumed}.

\kch{Throughout this} subsection we assume \kch{Model~$\Sigma$, which
is summarized in Table~\ref{tbl:Models} and uses} the \kch{disc}
\rw{and spiral potentials} described in \S\ref{sec:Model} with
parameter values given by Table~\ref{tbl:Values}, unless otherwise
stated. \rw{This model assumes an exponential radial velocity
dispersion profile, with scale length twice that of the assumed
exponential disc. We explore a range of normalisations at corotation, 
$\sigma_R(R=8$~kpc$) = \lbrace 5,~80\rbrace$~km~s$^{-1}$ \kch{in
$5$~km~s$^{-1}$ intervals}.}  \rw{The dependence of the typical (rms)}
amplitude of epicyclic excursions (equation~\ref{eqn:X}) \rw{on both
epicyclic frequency ($\kappa(R)$, equation~\ref{eqn:kappa}) and the
local value for the radial velocity dispersion ($\sigma_R (R)$) is
taken into account in the evaluation of the trapped fraction.}

\begin{figure}
\begin{center}
\includegraphics[scale=0.68]{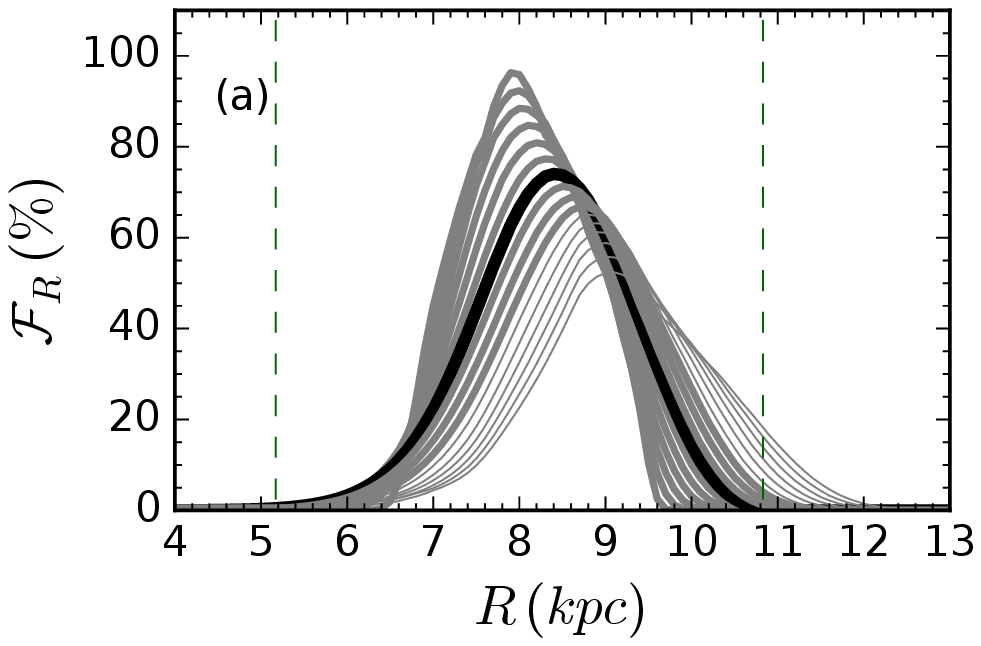}\\
\includegraphics[scale=0.68]{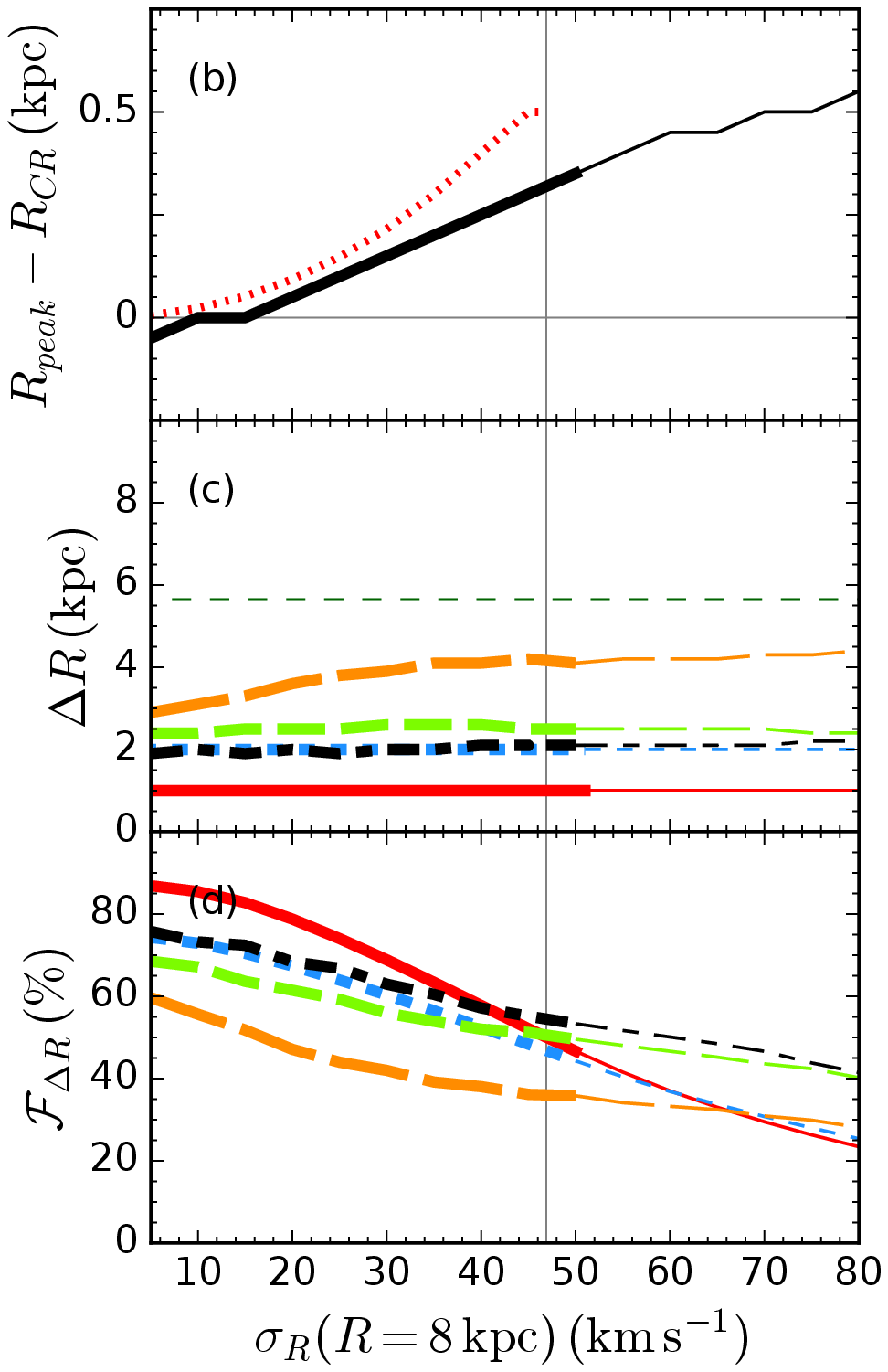}\\

\caption{Dependence of the integrated fraction on the normalisation of
the radial velocity dispersion profile \kch{in Model~$\Sigma$, where }
$\sigma_R\propto \Sigma(R)^{1/2} \propto e^{-R/R_\sigma}$ \kch{and
the} distribution \rw{function is} given by $f_G$.  \rwqq{\kch{T}he velocity dispersion is normalised \rw{at the radius of
corotation} such that $\sigma_R(R=8$~kpc$)=\lbrace
5,80$~km~s$^{-1}\rbrace$ in intervals of $5$~km~s$^{-1}$. } Panels show (a)
the radial distribution of the fraction of stars in trapped orbits
\kch{($\mathcal{F}_{R}$)}, \rwqq{with the curve for $\sigma_R(R=8$~kpc$)=35$~km~s$^{-1}$ in bold;} (b) the distance between the peak of the
radial trapped fraction profile and corotation ($R_{peak}-R_{CR}$)\rwqq{;}
(c) the radial range ($\Delta R$) within which the integrated fraction
is evaluated, and (d) the integrated fraction ($\mathcal{F}_{\Delta
R}$).  Line styles have the same meaning as in
\kch{F}igs.~\ref{fig:RMEII_CaptureDistribution_ADvsnoAD}-\kd{\ref{fig:RMEII_FitSigma_AD}}. \rwqq{Thin lines and segments of 
lines in all panels} indicate \kch{models}
that use \kch{a normalisation for the} radial velocity dispersion
$\sigma_R(R=8$~kpc$) > 50$~km~s$^{-1}$.
\kref{Thin, vertical, long-dashed (dark green) lines in panel~(a) mark the inner and outer Lindblad resonances.} }
\label{fig:RMEII_CaptureParameters_fS}
\end{center}
\end{figure}

The \kch{results} are illustrated in
Figure~\ref{fig:RMEII_CaptureParameters_fS} \kch{\rw{where individual
panels are analogues of}
Figures~\ref{fig:RMEII_CaptureDistribution_ADvsnoAD}-\kd{\ref{fig:RMEII_FitSigma_AD}})}
\rw{and show} (a) the radial distribution of the initial fraction of
stars in trapped \rw{orbits,} $\mathcal{F}_{R}(R)$\rw{;} (b) the
distance between the peak of $\mathcal{F}_{R}(R)$ and the radius of
corotation ($R_{peak}-R_{CR}$\rw{), as a function of velocity
dispersion normalisation;} (c) the various radial ranges ($\Delta R$)
within which the integrated fraction is evaluated, and (d) the
integrated fraction ($\mathcal{F}_{\Delta R}$).

\rw{The main difference between the curves for $\mathcal{F}_{R}$ in
Figure~\ref{fig:RMEII_CaptureParameters_fS}(a) and
Figures~\ref{fig:RMEII_CaptureDistribution_ADvsnoAD} is a steeper
decline at larger galactocentric radii in the case of the radially
varying velocity dispersion, particularly for larger values of the
normalisation at coorotation (8~kpc).  This reflects the fact that the
local velocity dispersion in the outer disc is lower compared to the
invariant case, so that the typical epicyclic excursions are lower,
such that stars observed in the outer disc are less likely to have
guiding centres within the capture region around corotation. The
trapped fraction is low in these regions, far from corotation, so the
differences are not large in terms of parameters such as the
integrated fraction, Figure~\ref{fig:RMEII_CaptureParameters_fS}(d). The widths of the various radial ranges (panel (c)) are also narrower, compared to the case of constant velocity dispersion, again reflecting the narrower distributions in panel (a).  Other than these minor quantitative differences, the qualitative results agree in showing a  decline of trapped fraction with increasing velocity dispersion. }

\subsection{Other Functional Dependencies for the Integrated Fraction}\label{sec:FncDepOfTrapFrac}

In this section, we assume the same model used in
\S\ref{sec:CapturedFractionSigNotRadDep}, where we adopt a Gaussian
velocity distribution \kch{at} each coordinate ($f_G$, described in
\S\ref{sec:fG})\kch{, radially fixed radial velocity dispersion
profile ($\sigma_R=35$~km~s$^{-1}$),} and exponential surface density
profile ($\Sigma(R)$, equation~\ref{eqn:SurfaceDensityProfile}).
Unless explicitly stated, model parameter values are given in
Table~\ref{tbl:Values}.

\begin{figure}
\begin{center}
\includegraphics[scale=1]{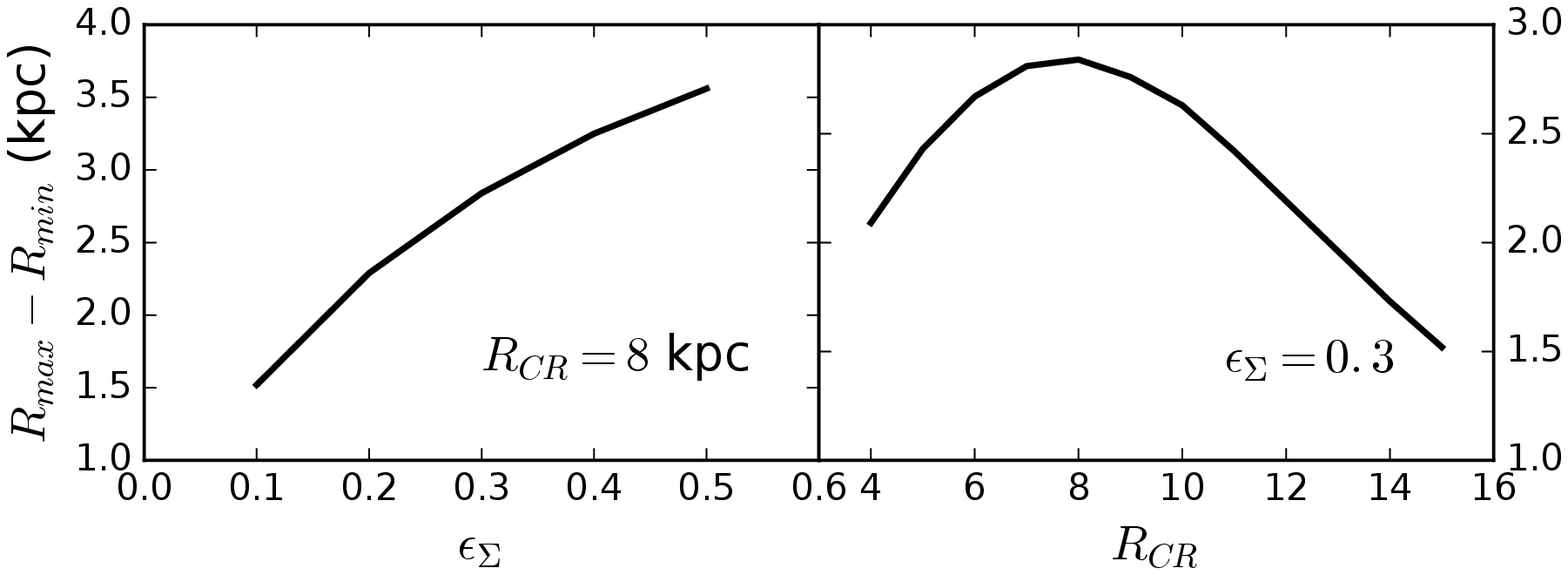}
\caption{\rwqq{Both curves represent} the maximum width of the capture region
($R_{max}-R_{min}$). The panel on the left shows the maximum size of the
capture region for $\epsilon_\Sigma=\lbrace 0.1,0.5\rbrace$ in intervals of
$0.1$ and fixed radius of corotation ($R_{CR}=8$~kpc).  The panel on the right shows the same measure for
varying radius of corotation, $R_{CR}=\lbrace 4,15\rbrace$~kpc in intervals of $1$~kpc, and fixed $\epsilon_\Sigma=0.3$.}
\label{fig:CapRegSize}
\end{center}
\end{figure}

The integrated fraction of stars initially in trapped orbits
($\mathcal{F}_{\Delta R}$) \kch{depends on the size and shape of the
capture region, which depends on the spiral pattern speed ($\Omega_p$)
and the underlying potential of the \rw{disc and spiral}
($\Phi(R,\phi)=\Phi_0(R)+\Phi_1(R,\phi)$) \citep[for a complete
description see][equation~17 and corresponding text]{DW15}.  For a
given pattern speed, the integrated fraction} increases with
increasing \rw{amplitude of the perturbation to the potential due to the spiral pattern ($|\Phi_s|_{CR}$), since the width} of the
capture region is larger for stronger spiral strengths (see
Appendix~\ref{sec:Capture}).  In our model, \rw{the strength of the spiral perturbation} 
(equation~\ref{eqn:SpiralAmplitude}) \rw{is written in terms of}  the radius of
corotation ($R_{CR}$), spiral pitch angle ($\theta$), number of spiral
arms ($m$), the \kch{assumed} surface density profile ($\Sigma(R)$),
and the fractional surface density \rw{enhancement} of the spiral potential
($\epsilon_\Sigma$) (see \S\ref{sec:Model}).  \rw{The model assumes a flat rotation curve of given circular velocity, so that `corotation radius' is a proxy for `pattern speed'.
Figure~\ref{fig:CapRegSize}  shows how  the maximum
width of the capture region ($R_{max}-R_{min}$) in our model varies with increasing  fractional surface density of the spiral pattern
($\epsilon_\Sigma$) for fixed corotation radius (left panel) and with increasing  radius of corotation ($R_{CR}$) for fixed fractional surface density (right panel)}. 

The curve in the lefthand panel of Figure~\ref{fig:CapRegSize} is
approximately a parabola, so that the (maximum) width of the capture
region scales as $\epsilon_\Sigma ^{1/2}$. For all other parameters
fixed (as here), $\epsilon_\Sigma$ is linearly proportional to the
amplitude of the spiral potential (equation~\ref{eqn:SpiralAmplitude}). Previous work \cite[their equation 12]{SB02},  established the expectation that the maximum radial range
of a trapped orbit would scale as the square root of the strength of
the imposed perturbation. The dependence seen in the figure then would
imply that the width of the capture region scale as the maximum radial
range of a trapped orbit. Since stars can be in trapped orbits while their physical coordinates lie outside the capture region, this correspondence \rwqq{is expected \kjd{to} be only approximate. }

\kref{The non-monotonic behaviour seen in the righthand panel reflects
how, assuming the same fractional surface density 
($\epsilon_\Sigma$), the radial range of the capture region does not
scale as $\sqrt{|\Phi_s|_{CR}}$ as one might naively expect.\footnote{\kref{Our 
prescription for the amplitude of the spiral pattern ($\Phi_s(R)$)
depends on the (exponential) surface density profile ($\Sigma(R)$) and the wave number ($k(R)$,
equation~\ref{eqn:WaveNumber}), so that the peak amplitude goes as
$|\Phi_s|_{CR}\propto R_{CR} \,e^{-R_{CR}/R_d}$ (see
\S\ref{sec:Model}) with maximum strength when $R_{CR}=R_d$.}  }
Figure~\ref{fig:RMEII_CaptureRegion} illustrates how, with increasing
spiral amplitude, not only does the radial range of the capture region broaden, but also  
its shape acquires a skew distortion from the simple sinusoidal model
assumed in the derivation of $\mathcal{F}_{\Delta
R}(\sigma_R\rightarrow 0)$ in \S\ref{sec:RadialDistributionSig}
\citep[the shape is mathematically described by the solutions to
equation~17 in][]{DW15}.  This skew becomes more prominent for
higher amplitude spiral perturbations.
Thus for lower values of the corotation radius, where the surface density profile ($\Sigma(R)$,
equation~\ref{eqn:SurfaceDensityProfile}) is exponentially greater, the resulting radial
range for the capture region is lower than the naive
expectation, where the least skew distortion per spiral strength appears to occur at $\sim 3R_d$
in our models.}

\rwqq{We turn next to investigations of how the trapped fraction depends on these parameters of the model potential.}

\subsubsection{\kch{Integrated F}raction \rw{as a Function of}  Spiral Strength}\label{sec:FracDepSpiralStrength}

\begin{figure}
\begin{center}
\includegraphics[scale=0.7]{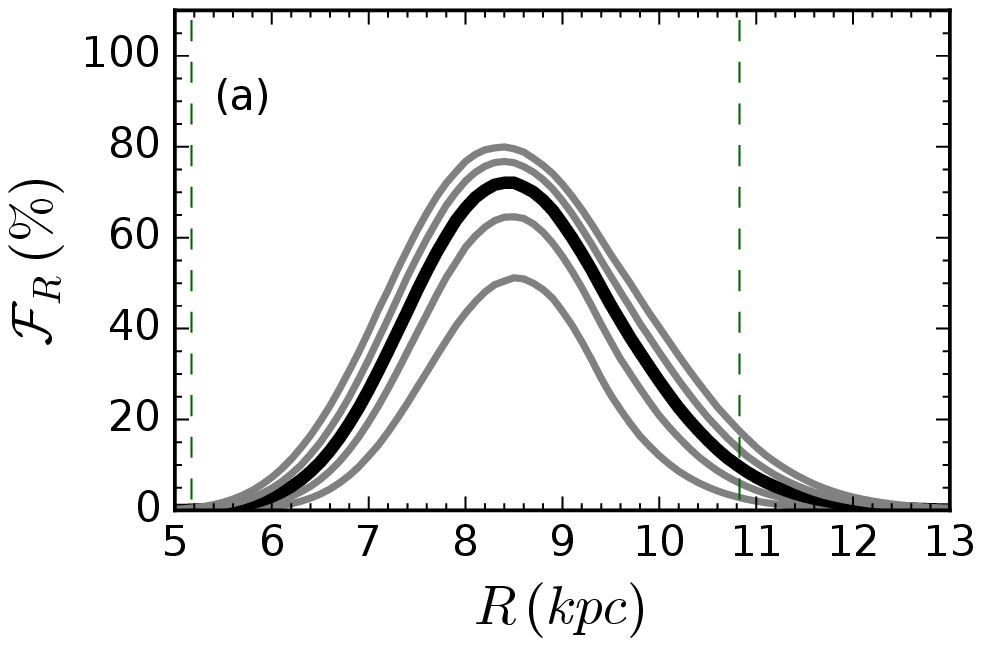}\\
\includegraphics[scale=0.7]{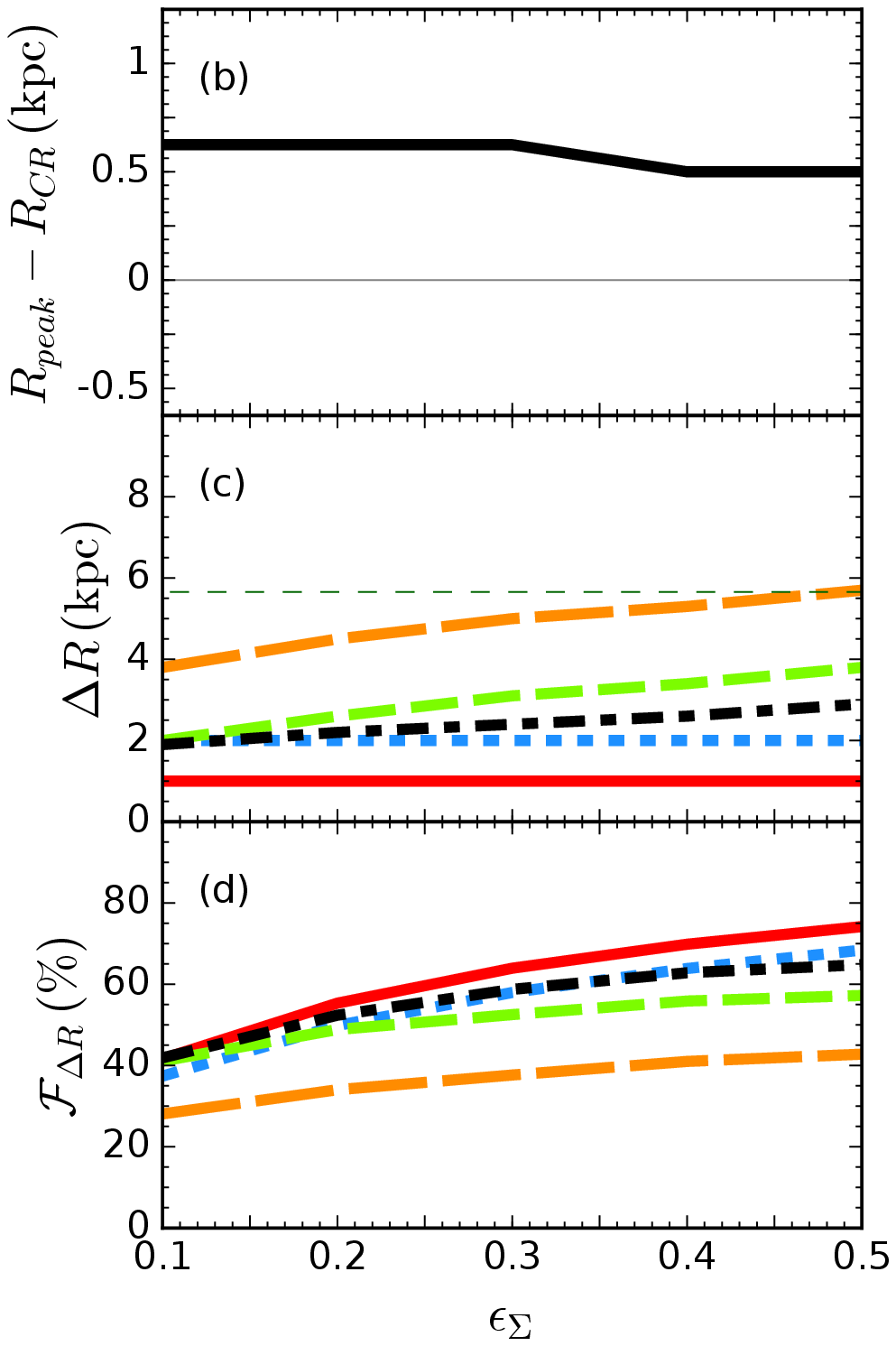}
\caption{Dependence of the integrated fraction \rwqq{of stars in
trapped orbits on the spiral strength, $\epsilon_\Sigma$, for
$\epsilon_\Sigma~=~0.1-0.5$ in intervals of 0.1. Otherwise the
underlying model assumptions are as in
Figure~\ref{fig:RMEII_CaptureDistribution_ADvsnoAD}, with
$\sigma_R=\kch{35}$~km~s$^{-1}$.  } Panels show (a) the radial
distribution of the fraction of stars in trapped orbits
\kch{($\mathcal{F}_{R}$)}, \rwqq{with the bold curve indicating
$\epsilon_\Sigma=0.3$;} (b) the \rwqq{offset} of the peak of the
radial trapped fraction profile from corotation
($R_{peak}-R_{CR}$)\rwqq{;} (c) the radial range ($\Delta R$) within
which the integrated fraction is evaluated, and (d) the integrated
\rw{fractions} of stars in trapped orbits ($\mathcal{F}_{\Delta R}$).
The line styles \rw{indicating the different definitions of the radial
range for integration} have the same \rw{meanings} as in
Figures~\ref{fig:RMEII_Offset_AD}-\kd{\ref{fig:RMEII_FitSigma_AD}}.}
\label{fig:RMEII_CaptureParameters_e}
\end{center}
\end{figure}

We here use the value for the fractional surface density
($\epsilon_\Sigma$) as a proxy for the amplitude of the spiral
potential at corotation ($|\Phi_s|_{CR}$) when all other parameters
are held constant.  In Figure~\ref{fig:RMEII_CaptureParameters_e}, we
use the same five values for the fractional surface density
($\epsilon_\Sigma$) as in Figure~\ref{fig:CapRegSize} and a radially
\rw{invariant} radial velocity dispersion
(\kch{$\sigma_R=35$}~km~s$^{-1}$).  \rw{Panel~(d) of
Figure~\ref{fig:RMEII_CaptureParameters_e} \rw{shows} that the
integrated fraction of stars in trapped orbits ($\mathcal{F}_{\Delta
R}$) increases \kref{smoothly} with \rw{increasing} fractional
surface density ($\epsilon_\Sigma$).  The shape of the curves in
Figure~\ref{fig:RMEII_CaptureParameters_e}(d) are remarkably similar
to shape of the curve in the left hand side of
Figure~\ref{fig:CapRegSize}, indicating 
that the integrated fraction scales with the the maximum width of the
capture region \rw{(or area, for given value of $R_{CR}$)}.  }  A
similar \rw{behaviour is expected for the dependence of the integrated fraction
on any other parameter that enters linearly in the expression for
spiral strength (equation~\ref{eqn:SpiralAmplitude}), such as number
of arms ($m$).}  

\kref{There is a $0.1$~kpc decrease in the offset between corotation and the peak
of the radial trapped fraction profile from $\epsilon_\Sigma=0.3$ to $\epsilon_\Sigma=0.4$,
as shown in panel~(b) of Figure~\ref{fig:RMEII_CaptureParameters_e}.  This could arise from
the relative importance of the skew in the shape of the capture region with increasing amplitude
for the perturbing potential (see discussion in \S\ref{sec:FncDepOfTrapFrac}).
However, it is difficult to draw significant conclusions about this shift since changes in the 
values for $R_{peak}-R_{CR}$ between models are less than or on order of $\delta R$ (Table~\ref{tbl:Values}).}

There is a clear trend that the integrated fraction of stars in trapped orbits ($\mathcal{F}_{\Delta R}$) increases with increasing spiral amplitude at corotation ($|\Phi_s|_{CR}$).  The fitting coefficients in equation~\ref{eqn:CapturedFractionLinearFit} ($\xi_{\mu,\beta}$) also depend on spiral strength, but \rw{we have found no simple scaling or transformation}. 

\subsubsection{Integrated \rw{Trapped} Fraction \rw{as a Function of} \kch{Pattern Speed}}\label{sec:FracDepCR}

In \S\ref{sec:FracDepSpiralStrength} we demonstrated that the
integrated fraction of stars in trapped orbits ($\mathcal{F}_{\Delta
R}$) depends on the \rw{amplitude of the spiral perturbation to the potential} at  corotation ($|\Phi_s|_{CR}$)
since the maximum \rw{width}  of the capture region increases with increasing
\rw{strength of the spiral perturbation}.  \kch{Our prescription for the amplitude of the
spiral pattern, ($\Phi_s(R)$, equation~\ref{eqn:SpiralAmplitude})
depends} on the \rw{both the disc  surface density profile, which we have chosen to be an exponential} 
(\kch{$\Sigma(R)$,} equation~\ref{eqn:SurfaceDensityProfile}) and the
wave number (\kch{$k(R)$,} equation~\ref{eqn:WaveNumber}), \rw{such that}  the
peak amplitude \rw{scales} as $|\Phi_s|_{CR}\propto R_{CR} \,e^{-R_{CR}/R_d}$
(see \S\ref{sec:Model}).  \kch{In the introduction to this section
(\S\ref{sec:FncDepOfTrapFrac}) and the} right-hand panel of
Figure~\ref{fig:CapRegSize} we showed that the maximum width of the
capture region ($R_{max}-R_{min}$) scales \rw {non-monotonically} with the
radius of corotation \kch{.  In this subsection we explore how and why
the integrated fraction depends on the radius of corotation
($R_{CR}$), and its corresponding spiral pattern speed ($\Omega_p =
v_c/R_{CR}$), including the size of the capture region.}

\begin{figure}
\begin{center}
\includegraphics[scale=0.7]{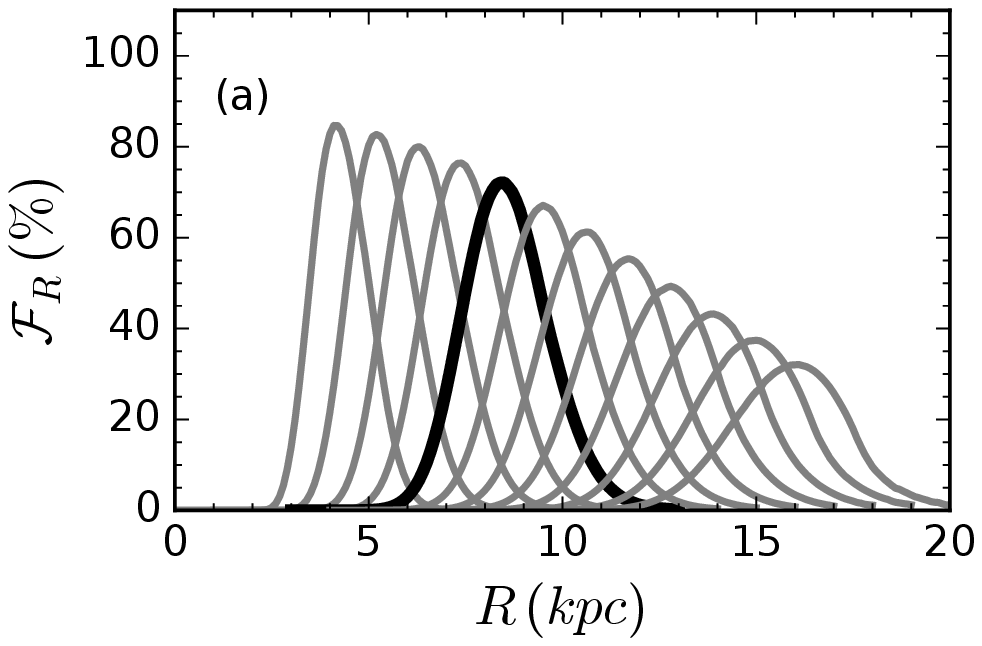}\\
\includegraphics[scale=0.7]{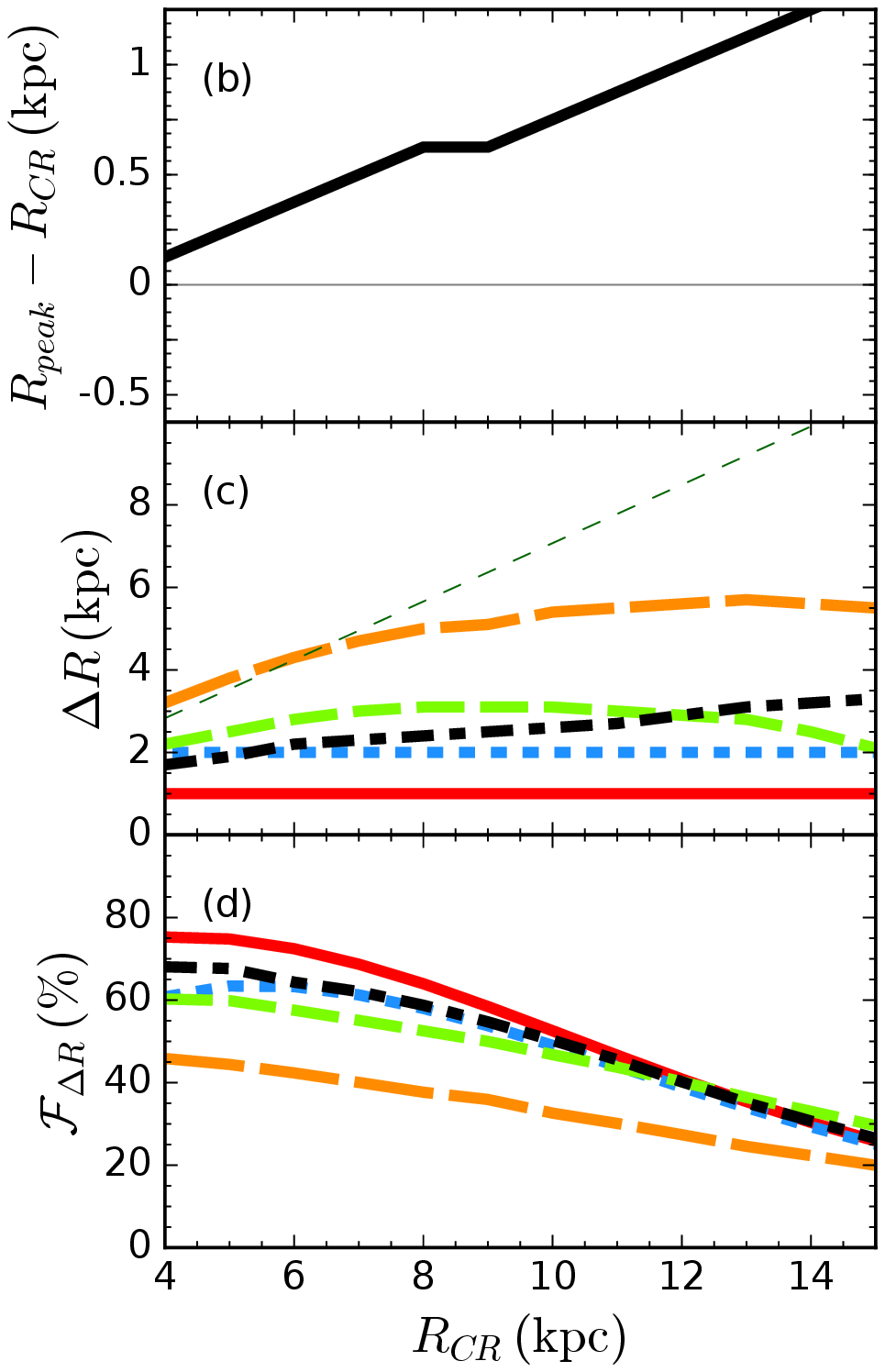}
\caption{Dependence of the integrated fraction on the radius of
corotation in the \rwqq{current} adopted \rwqq{model, for $R_{CR}~=~4-15$~kpc in 1~kpc intervals. Remaining model assumptions are as 
in }
Figure~\ref{fig:RMEII_CaptureDistribution_ADvsnoAD} with
$\sigma_R=\kch{35}$~km~s$^{-1}$.  
Panels show (a)
the radial distribution of the fraction of stars in trapped orbits
\kch{($\mathcal{F}_{R}$)}, \rwqq{with the bold curve for $R_{CR}=8$~kpc;} (b) the \rwqq{offset of  the location of the peak value of the trapped fraction}  from corotation ($R_{peak}-R_{CR}$)\rwqq{;} (c) the
radial range ($\Delta R$) within which the integrated fraction is
evaluated, and (d) the integrated fraction of stars in trapped orbits
($\mathcal{F}_{\Delta R}$)  Line styles are the same as in
Figures~\ref{fig:RMEII_Offset_AD}-\kd{\ref{fig:RMEII_FitSigma_AD}}.}
\label{fig:RMEII_CaptureParameters_CR}
\end{center}
\end{figure}

\kch{The size of epicyclic excursions, as well as the position and shape of the capture region, affect the \rw{radial} distribution of stars in trapped orbits ($\mathcal{F}_R$).}
In Figure~\ref{fig:RMEII_CaptureParameters_CR}, we set $\sigma_R=\kch{35}$~km~s$^{-1}$ at all radii\kch{, let} the phase space distribution \kch{be} given by $f_G(\mathbf{x},\mathbf{v})$ (\S\ref{sec:fG})\kch{, and calculate \rw{the various measures of trapped fraction ($\mathcal{F}_R$)} for pattern speeds corresponding to radii of corotation between $4-15$~\rw{kpc,} in $1$~kpc intervals}.

For fixed radial velocity dispersion ($\sigma_R$) \rw{and flat
rotation curve,} the \rw{typical stellar radial epicyclic excursions
increase linearly with galactocentric radius}, $X\propto R$ (see
equation~\ref{eqn:X}).  \kch{The distribution of stars in trapped
orbits ($\mathcal{F}_{R}$) therefore depends on the radius of
corotation ($R_{CR}$) since \rw{both} the maximum amplitude of epicyclic
excursions ($X$) and the width of the capture region
($|R_{max}-R_{CR}|\approx |R_{CR}-R_{min}|$) set the maximum distance
from corotation for stars in trapped orbits \rw{and $R_{CR}$ sets the area of the capture region} (see discussion following
equation~\ref{eqn:X}).  The maximum range \rw{in radial coordinate} for stars in trapped orbits
is illustrated in panel~(a) of
Figure~\ref{fig:RMEII_CaptureParameters_CR} and best quantified by
$\Delta R_{5\%}$ (long-dashed, orange) in panel~(c).  The shape \rw{of the curve for}
$\Delta R_{5\%}$ in Figure~\ref{fig:RMEII_CaptureParameters_CR}(c) is
consistent with \rw{that for the width} of the capture region as a function of
corotation (\rw{see Figure~\ref{fig:CapRegSize}}) \rw{combined with the} linear increase in \rw{typical}  maximum radial \rw{epicyclic excursions
($X\propto R$).}} 

Equation~\ref{eqn:RPeakPredict} predicts that the value for $R_{peak}-R_{CR}$ scales as $R_{CR}$ for fixed velocity dispersion (a consequence of epicyclic excursions scaling with radius in this regime).  Panel~(b) of Figure~\ref{fig:RMEII_CaptureParameters_CR} illustrates this point.  \kch{T}he curve in this plot is not smooth (\kch{and also} in panel~(b) of Figure~\ref{fig:RMEII_CaptureParameters_e}).  \kref{As stated in \S\ref{sec:FracDepSpiralStrength}, this could be due to the relative importance of the
skew in shape of the capture region with increasing spiral amplitude but it is unclear} since changes in 
the values for $R_{peak}-R_{CR}$ between \kch{models} are less than or on order of \rw{$\delta R$} (Table~\ref{tbl:Values}).

\kch{The trend is that the integrated fraction ($\mathcal{F}_{\Delta
R}$) generally decreases with increasing radius of \rw{corotation, equivalent to decreasing pattern speed.}.  All
evaluations of $\mathcal{F}_{\Delta R}$ will naturally decrease \rw{in terms of the maximum value,} since
$\mathcal{F}_{R}$ broadens \rw{with}
increasing $R_{CR}$ (\rw{as also discussed}  in
\S\ref{sec:CapturedFractionSigNotRadDep}). \rw{Should the} offset
($R_{peak}-R_{CR}$) be large enough \rw{for the peak to be beyond the annulus \rw{over which the integrated fraction is evaluated, as can happen for fixed annular widths such as $\Delta
R_{1,2}$,} the integrated fraction in that case ($\mathcal{F}_{\Delta R}$) decreases
more rapidly with increasing corotation radius than when
$\mathcal{F}_{\Delta R}$ is evaluated using non-fixed radial ranges} 
(i.e.~$\Delta R_{5\%,25\%,FWHM}$).}

\section{\kch{I}ntegrated \rw{Trapped} \kch{F}raction \kch{in Models of the Galactic Disc}}\label{sec:CapFracInModels}

\kch{In \S\ref{sec:CapturedFraction}, we \rw{established} that the
value of the integrated fraction \rwqq{depends strongly  on the} \rw{value of the} radial
velocity dispersion at corotation ($\sigma_R(R_{CR})$). 
We also \rw{found}  that the integrated
fraction depends on the \rw{area and location} of the capture region, the size
of epicyclic excursions, and the spiral strength and pattern speed.
These analyses in hand, we are equipped to interpret how \rwqq{our adopted forms for the stellar} distribution function, \rwqq{including 
particular prescriptions} for the radially dependent radial velocity
dispersion profile\kdd{,} \rwqq{affect the derived behaviours} of the integrated fraction
over the disc.}

\begin{figure}
\begin{center}
\includegraphics[scale=0.7]{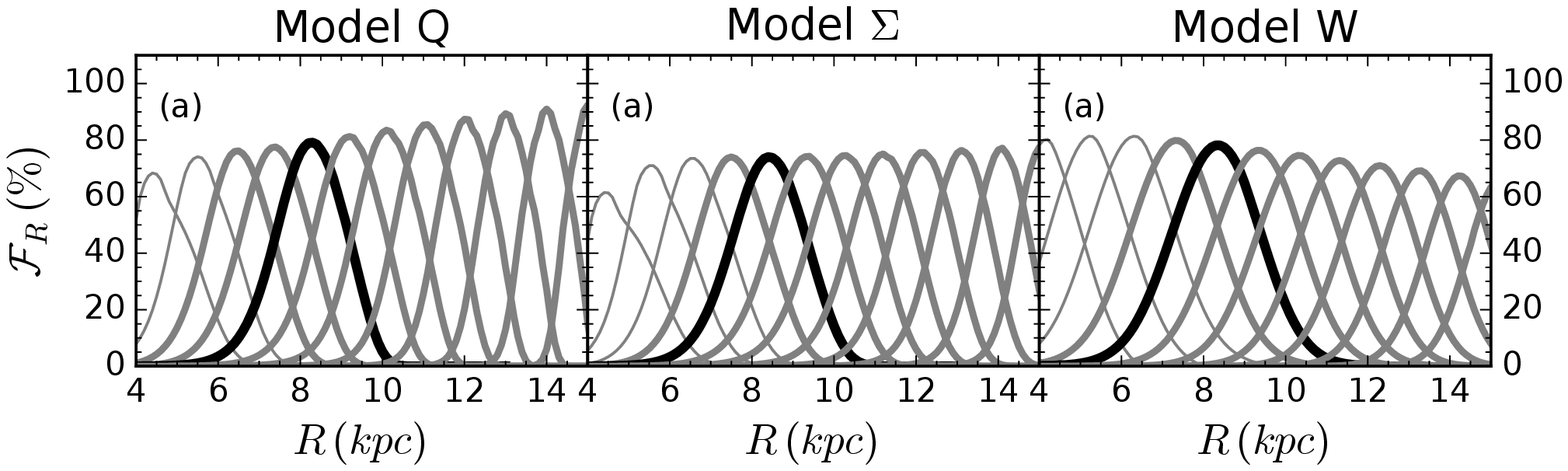}\\
\includegraphics[scale=0.7]{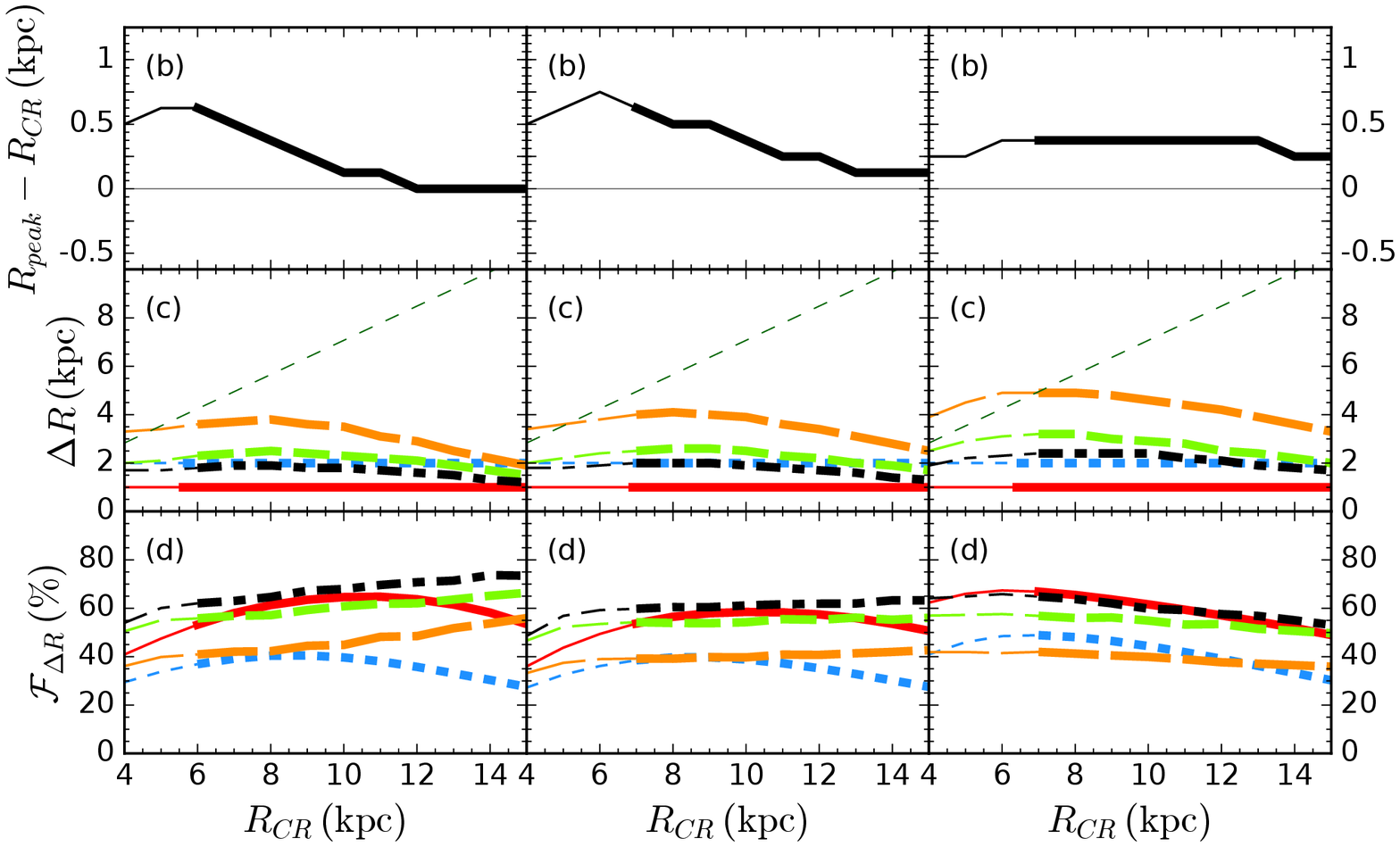}\\
\caption{\rwqq{As Figure~\ref{fig:RMEII_CaptureParameters_CR}, 
the dependence of the integrated fraction on the radius of
corotation, now for} \rw{the three classes of model} in Table~\ref{tbl:Models}.  Each model
assumes \rw{that the radial velocity dispersion profile is radially dependent, with a different form of the profile in each class of model}.
Panels show (a) the radial distribution of the fraction of stars in
trapped orbits \kch{($\mathcal{F}_{R}$)}\rwqq{;} (b) the \rwqq{offset of the location of the peak value of the trapped fraction from the} radius of
corotation ($R_{peak}-R_{CR}$)\rwqq{;} (c) the radial range ($\Delta R$)
within which the integrated fraction is evaluated, and (d) the
integrated ($\mathcal{F}_{\Delta R}$).  Line styles have the same
meaning as in
\kch{Figs.}~\ref{fig:RMEII_CaptureDistribution_ADvsnoAD}-\kd{\ref{fig:RMEII_FitSigma_AD}}.
\rwqq{Thin lines and line segments  indicate that the value of the radial velocity dispersion at $R_{CR}$ exceeds}  
50~km~s$^{-1}$.}
\label{fig:RMEII_CaptureParameters_sigexp}
\end{center}
\end{figure}

\rwqq{Our results for the three \kdd{fiducial} models outlined in
Table~\ref{tbl:Models} are shown in}
Figure~\ref{fig:RMEII_CaptureParameters_sigexp}, \rwqq{for different
values of the corotation radius such that $R_{CR}$ varies between
$4-15$~kpc,} in $1$~kpc intervals.  Each model adopts the parameter
values in Table~\ref{tbl:Values}\kdd{, where the} normalisation of the
radial velocity dispersion profile ($\sigma_{R}(R)$) \kdd{for each
model is given in Table~\ref{tbl:Models}}.  \kdd{In this section, we
will only consider regions of the disc where $\sigma_R \leq
50$~km~s$^{-1}$.}

\rwqq{For all values of the corotation radius, we find that the}
trends in the \rw{width} of the unconstrained radial ranges for
evaluation ($\Delta R_{5\%}$, $\Delta R_{25\%}$, and $\Delta
R_{FWHM}$) and their corresponding integrated fractions
($\mathcal{F}_{5\%}$, $\mathcal{F}_{25\%}$, and $\mathcal{F}_{FWHM}$),
primarily depend on \rwqq{both} the \kdd{area} of the capture region
\kdd{and the \rwqq{typical} amplitude of epicyclic excursions
\rwqq{for} stars in trapped orbits \rwqq{(parameterised by the rms
maximum amplitude, $X$)}.  \rwqq{These in turn depend} on the value of
$R_{CR}$.  \kdd{The maximum radial width of the capture region, which
closely scales in amplitude with its area, is shown as a function of
corotation in the right hand panel of} Figure~\ref{fig:CapRegSize}.}
The width of the capture region \rwqq{depends on the adopted
underlying potential} \kdd{\citep{DW15}}.  \kdd{The curve in the right
hand} panel of Figure~\ref{fig:CapRegSize} is \kdd{\rwqq{applicable to
all of our} fiducial models and is} consistent with the maximum radial
width for the capture region roughly scaling with the radius of
corotation as $(R_{max}-R_{min})\propto R_{CR}\,e^{-R_{CR}/3R_d}$.
\kdd{The area of }the capture region \rwqq{influences} the
\rwqq{radial width of} the distribution of stars in trapped
\rwqq{orbits and,} more importantly, the \rw{local surface density or}
\textit{number} of stars in trapped orbits.  \kdd{Taking into account
that the shape and area of the capture region for a given radius of
corotation is the same for all fiducial models \rwqq{(assuming fixed
remaining parameters of the spiral) since they assume the same
underlying potential}, we can explore how the kinematics of each model
affects the integrated fraction of stars in trapped orbits.}

\newrw{The radial dependence of the velocity dispersion affects} the integrated 
fraction in two ways.  First, the radial velocity dispersion sets the amplitude of radial
epicyclic excursions, $X\propto\sigma_R(R)/\kappa(R)$
(equation~\ref{eqn:X}).  In turn, the sizes of epicyclic excursions affects the radial width of
the distribution of stars in trapped orbits ($\mathcal{F}_{R}$) since these determine the typical difference between a star's physical radial coordinate and that of its guiding centre.
Second, the velocity dispersion sets the amplitude for the offset $R_{peak}-R_{CR}$ 
(panels~(b))
through its dependence on the asymmetric drift (see
eqn~\ref{eqn:RPeakPredict} earlier). 
\newrw{The behavior of the  curves for each model in
Figure~\ref{fig:RMEII_LzR}, which illustrate how the velocity dispersion underlies the shape of the 
orbital angular momentum distribution, is related to this interdependence.  The angular momenta} decrease toward the galactic
centre, compared to the case of all stars on circular orbits, as velocity dispersion increases.  The overall consequence is that a lower fraction of stars near the galactic centre meets the capture
criterion for each model.
The values for unconstrained measures of the integrated fraction 
($\mathcal{F}_{\Delta R}=\mathcal{F}_{5\%,25\%,FWHM}$) quantify the shape of the distribution
of stars in trapped orbits, where the range in radii
for \newrw{the various measures} $\mathcal{F}_{R}$ is
closely approximated by that given by $\Delta R_{5\%}$
(long-dashed, orange curve in panels~(c)). 
The assumed radial velocity dispersion profile ($\sigma_R(R)$) is 
therefore integral to understanding the integrated
fraction $\mathcal{F}_{\Delta R}$  for each model since it sets \newrw{the} 
radial dependence \newrw{of the radial offset, $R_{peak}-R_{CR}$, and the shape (measured} by $\Delta R$) of the distribution of stars in trapped orbits.  
The following is a more detailed exploration of the distributions shown in Figure~\ref{fig:RMEII_CaptureParameters_sigexp} for each model.

The radial velocity dispersion profile for Model~W
($\sigma_R(R)\propto e^{-R/3R_d}$) is relatively shallow
(Figure~\ref{fig:RMEII_sigR_exp_vs_Q}, dot-dashed, black) and leads to
\rwqq{an approximately} constant value for $R_{peak}-R_{CR}$ as
the value for the corotation varies 
(Figure~\ref{fig:RMEII_CaptureParameters_sigexp}(b), last column)\kjd{.  
The} curve for $\Delta R_{5\%}$ peaks at approximately $3 R_d$ where its value is nearly
2~kpc greater than its minimum value.  The shape of
this curve \rwqq{arises from two competing effects.}  The first
is the \rwqq{dependence of the} area of the capture region \rwqq{on
the value of $R_{CR}$,} as described above and in the introduction to
\S\ref{sec:FncDepOfTrapFrac}.  Second, the \rwqq{amplitude} of radial
epicyclic excursions \kdd{in Model~W} goes as
\kjd{$X_W\propto R\kdd{\,e}^{-R/3R_d}$.}
\kdd{Thus, in Model~W, the} \rwqq{variation of the width of}
$\mathcal{F}_{R}$ \kdd{from epicyclic orbits} equals the radial
dependence for the total number of stars in trapped orbits \rw{and
these two effects \rwqq{act in opposing
fashions}}.  The final \kdd{result of} this balance is that the value
of the integrated fraction for $\mathcal{F}_{5\%}$ (and other
unconstrained measures of $\mathcal{F}_{\Delta R}$) is nearly constant
with radius.  Fixed measures for $\mathcal{F}_{\Delta R}$
(i.e.~$\mathcal{F}_{1,2}$) \rwqq{follow} $(R_{max}-R_{min})$ and \kdd{do}
not have a significant radial dependence on asymmetric drift.

The radial velocity dispersion profile for Model~Q approaches zero at
large radii.  Thus with increasing \rwqq{values for the corotation}
radius, the \rw{offset of} the peak of the distribution of the trapped
fraction ($R_{peak}-R_{CR}$) goes to \rwqq{zero. Further,} the
unconstrained radial ranges converge to the \rwqq{width} of the
capture region (as most of the stars in trapped orbits are
\kch{physically} in the capture region)\kdd{, the} \kch{maximum}
values for $\mathcal{F}_{R}$ approach $100\%$ (as most of the stars
\kch{near corotation} are in trapped orbits)\kdd{, and
$\mathcal{F}_{5\%}$ approaches $\sim 64\%$ \rwqq{(the result from
equation~(16) in section 3.1.1 earlier)}.}  \kdd{The \rwqq{amplitude}
of radial epicyclic excursions for Model~Q goes as $X_Q\propto
R^2\,e^{-R/R_d}$, \rwqq{which peaks} closer to the galactic centre
(at $2R_d$) than \rwqq{the expression} for the radial range of the
capture region at ($\sim 3R_d$).  Should the offset $R_{peak}-R_{CR}$ be
constant, this would have the effect of broadening the distribution of
the trapped fraction compared to Model~W.  However, the offset
$R_{peak}-R_{CR}$ increases with decreasing corotation \rwqq{radii, reaching  a value of} nearly
half the width of the capture \kjd{region.  
The overall consequence is that
a lower fraction of stars near the galactic centre meets the capture
criterion, compared to Model~W,} and the integrated fraction for Model~Q
thus decreases toward the galactic centre.}

\kjd{The trends in the curves for Model~$\Sigma$ (Figure~\ref{fig:RMEII_CaptureParameters_sigexp}, middle column) are similar to those for
Model~Q.}  \kdd{\kjd{Both these models assume the same form for the distribution function ($f_G$), but different radial velocity dispersion profiles.}  The normalisation for the radial velocity dispersion
profile for Model~$\Sigma$ sets the local value for $\sigma_R$ to
greater values than in Model~Q at all radii.  Model~$\Sigma$ also has the
steepest \rwqq{(declining with radius)} radial velocity dispersion profile of} all models and
consequently \rwqq{has} the lowest fraction of stars that meet the capture
criterion toward the galactic centre (see also the dashed, blue curve
in Figure~\ref{fig:RMEII_LzR}).  The radial velocity dispersion
profile for Model~Q was selected in order to approximate a disk that
could support \rwqq{gravitational} instabilities at all radii.  Interestingly, this also \rwqq{produces} a higher
integrated fraction at all corotation radii.

For all models with radially dependent radial velocity dispersion
\rwqq{profiles,} the value for the integrated fraction
($\mathcal{F}_{\Delta R}$) \rwqq{for any given radial range for
evaluation ($\Delta R$) is} rather constant ($\pm \sim 10\%$) as a
function of corotation radius.  
\kref{This nearly constant value for the integrated fraction
primarily arises from (1) the identical radial dependences for the
radial range of the capture region (Figure~\ref{fig:CapRegSize}) for each model and (2) the similar distribution of 
orbital angular momenta for stars with guiding centres within the capture region -- where
 the associated measurables are the size of epicyclic 
excursions ($X(R)$) and the offset offset $R_{peak}-R_{CR}$ -- due to the 
assumed kinematics for each model.}
As shown in \kref{\S\ref{sec:CapturedFractionSigRadDep}},
the integrated fraction decreases with increasing value for the
normalisation of the velocity dispersion profile.

\section{Discussion}\label{sec:Discussion}

\subsection{\kch{Integrated} \rw{Trapped} Fraction as an Upper Limit}\label{sec:UpperLimit}

\rwqq{The initial distribution of stars in trapped orbits derived
above should be interpreted as an upper limit, for the following
reasons.}  First, \kch{the parameter} $\Lambda_{nc,2}(t)$\kch{, used
to evaluate whether or not a star is in a trapped orbit,} is
explicitly a time-dependent quantity (equation~\ref{eqn:Lambda_nc2}) \rwqq{and we showed in Paper~I} that stars which initially \kch{satisfy} the
capture criterion \kch{could be} scattered\footnote{We define \lq\lq
scattering" as an event where a star' s change in orbital angular
momentum is correlated with a change in random orbital energy,
distinct from radial \kch{migration}.} out of a trapped orbit before \rwqq{the torques from the spiral perturbation had time to change significantly that star's} orbital angular momentum.
Scattering events could affect the fraction of stars in trapped orbits
over time (see discussion in Paper~I, \S3.3).  This is an especially
important point for stars in trapped orbits that \kch{either (1) have}
large radial excursions from the guiding centre, since such stars
sample a larger fraction of the underlying \kch{disc} potential and
are therefore are more likely to scatter as they encounter
inhomogeneities in the \kch{disc},\footnote{A star in a trapped orbit
that has large excursions from its guiding centre
\kch{could} have a coordinate position that is distant from the capture region.  In a realistic \kch{disc}, such a trajectory would likely lead to scattering via interactions with inhomogeneities in the \kch{disc} potential that are associated with objects such as dark matter substructure and giant molecular clouds.  When modelling radial \kch{migration} in the \kch{disc}, one should also take care that the choice for $\Delta R$ is less than the radial distance between spiral arms (here $\lambda(R)=2\pi/k(R)$), where the radial wavenumber $k(R)$ (equation~\ref{eqn:WaveNumber}) is proportional to the number of spiral arms ($m$).} or \kch{(2) meet a resonant criterion in addition to the corotation resonance -- e.g.~ the path for the guiding centre radius crosses an ultra-harmonic Lindblad resonance}. 

Indeed, \cite{BLJ13} \rwqq{observed that a sufficiently long-lived
spiral perturbation caused not only the pattern of angular momentum
changes that correspond to trapped `horseshoe' orbits, but also a later pattern of opposite angular momentum changes for
stars close to the corotation radius, consistent with scattering
processes.}

\kref{Although  it is generally accepted that discs are kinematically heated over time, the mechanisms and their relative importance aren't entirely clear.}
\cite{CS85} \kref{argued} that a rapidly growing spiral perturbation would cause a non-adiabatic response in the stellar \kch{disc} over a broad radial range and a second order change in the orbital angular momentum distribution.  
\kref{In a high resolution follow up study using a pure stellar disk, \cite{Fujii11} showed that even long lived material spiral arms  kinematically heat the disk and they further supported this claim with analytic arguments.  Indeed, \citeauthor{Fujii11} based their analytic argument for  heating rates on those from
GMCs and other small scale fluctuations in the galactic potential \citep{SS53,Wielen77,Lacey84}.}
\kref{In any case, we have shown that} a \kch{kinematically} heated population of stars has a lower fraction of stars in trapped orbits.  \kch{This implies} that such heating by the spiral would \kch{over time} lead to fewer stars being captured in trapped orbits than would be expected from the initial integrated fraction.

\kch{Our} analysis assumes that all orbits are constrained to a 2D
\kch{disc}.  This approximation is valid for stellar orbits that are
nearly circular since the vertical and radial actions are separable
\citep[this is an underlying assumption in the derivation of the
capture criterion,][]{DW15}.\footnote{Indeed, we expect the capture
criterion to be less robust for stars with highly non-circular
orbits.}  \kch{The vertical and radial actions \rw{may be} coupled}
(and thus the validity of the capture criterion is untested) for a
star that has large random motions.  \rwqq{Further, the
azimuthal force from the spiral (responsible for the changes in orbital angular momentum of a trapped star) will be weaker above/below the mid-plane. Our 
neglect of  vertical motions thus overestimates how strongly stars
in 3D orbits would} interact with the spiral pattern.  We therefore expect
there is a systematic overestimate of the initial fraction of stars in
trapped orbits with increasing velocity dispersion.  As mentioned in
\S\ref{sec:PreviousWork}, \cite{SSS12} found an exponential decrease
in the RMS change in orbital angular momentum for an ensemble of stars
due to a transient spiral with increasing scale height of the
population.

We defer a more thorough discussion of the time-dependence of the fraction of stars captured in trapped orbits to a later paper in this series (Paper~IV).

\subsection{Comparison to \kch{P}revious \kch{Theoretical W}ork}\label{sec:PreviousWork}

\kch{A definition for maximum} efficiency of radial \kch{migration from} a single spiral pattern is the case \kch{in which}  the RMS change in orbital angular momentum for the ensemble of stars in trapped orbits, $\langle(\Delta L_z)^2\rangle^{1/2}$, is maximized over the spiral lifetime.  \kch{This definition depends on the fraction of stars in trapped orbits (evaluated using $\mathcal{F}_{\Delta R}$ in this paper), the amplitude of the change in \textit{each} trapped star's angular momentum (maximum values can be estimated using the radial width of the capture region - $R_{max}-R_{min}$), and the degree of scattering out of trapped orbits on timescales less than the timescale for radial migration.} 
  
\cite{SSS12} used a simulation to study the relationship between $\langle(\Delta L_z)^2\rangle^{1/2}$ for a population of stars and that population's scale height \kch{or} initial radial velocity dispersion.  \kch{They} used seven \kch{tracer} populations of test particles in \kch{a 3D \kch{disc} with gravitational field defined by tapered, exponential thin and thick \kch{disc}s and} two spiral arms\kch{.  In two separate experiments} each population \kch{was assigned either} a different scale height \kch{or} initial radial velocity dispersion\kch{.} 
They found that the value for $\langle(\Delta L_z)^2\rangle^{1/2}$  for a given stellar population decreases exponentially with linearly increasing values for the scale height.  They also found that $\langle(\Delta L_z)^2\rangle^{1/2}$ is smaller for populations with higher initial radial velocity dispersions.\footnote{\citet{SSS12} do not find a functional form for the relationship between the RMS change in orbital angular momentum ($\langle(\Delta L_z)^2\rangle^{1/2}$) and initial radial velocity dispersion ($\sigma_{r,0}$).  However, we find that the closest fit to the data \kch{goes} as $\langle(\Delta L_z)^2\rangle^{1/2} \propto \sigma_{r,0}^{1/2}$.} 
It is encouraging that we \kch{also find a} decline in the value for the integrated fraction of stars in trapped orbits, $\mathcal{F}_{\Delta R}$, with higher values for the radial velocity dispersion (\S\ref{sec:IntegratedFractionSigRConstant}), strengthening the argument that the efficiency of radial \kch{migration} depends on the radial velocity dispersion.

It is not obvious how to further compare our results to the trend
found \rwqq{by \citeauthor{SSS12}}  The value of $\langle(\Delta
L_z)^2\rangle^{1/2}$ is, in general, a time dependent quantity (\rwqq{for example, due to} the effects of scattering) for the entire population of
stars.  \cite{SSS12} measured $\langle(\Delta L_z)^2\rangle^{1/2}$
over the lifetime of the simulation in order to evaluate the final
$\langle(\Delta L_z)^2\rangle^{1/2}$ induced by a transient spiral
pattern from initial growth through final decay.  The integrated
fraction ($\mathcal{F}_{\Delta R}$) \rwqq{we have derived in this paper} is a measure only of the initial
fraction of stars that \textit{may} migrate radially.  We therefore
interpret the fall off in the initial fraction of stars in trapped
orbits with increasing velocity dispersion as a first step toward
isolating the physics important to understanding the time-dependent
value of $\langle(\Delta L_z)^2\rangle^{1/2}$.  The time dependence
for the fraction of stars in trapped orbits \rwqq{obviously} depends on the rate of
\rwqq{scattering. Nonetheless,} it is expected that the
effects of scattering will be less important for \rwqq{more open} spiral
arms (since even a star with large radial excursions from its guiding
centre will not interact with the spiral away from corotation) and for
\rwqq{perturbations with} fewer spiral arms ($m=2$ rather than $m=4$ -- since the Lindblad
resonance will be farther from the radius of
corotation).\footnote{Figure~7 and Table~3 from \cite{SSS12} show that
the value for $\langle(\Delta L_z)^2\rangle^{1/2}$ is smaller for
higher number of spiral arms, even when the prescription for the
initial radial velocity dispersion is smaller for higher number of
spiral arms (Table~1).}

Further, we cannot make any predictions about whether or not the vertical position or motion of a star affects its trapped status because we have assumed a 2D \kch{disc} on the premise that vertical action and radial action are separable (see Paper~I, \S 2).  Studies of simulated \kch{disc} galaxies support the notion that the vertical velocity dispersion of a population \kch{n}egatively affects the degree to which it migrates radially \citep{VC14,VCdON16,Grand16}.
However, should the ratio $\sigma_R/\sigma_z$ be \kch{approximately} constant, we expect a similar trend in the initial fraction of stars in trapped orbits as in equation~\ref{eqn:CapturedFractionLinearFit}.

\citet[][their equations~12]{SB02} predict that the maximum change in
angular momentum for a star that migrates radially, $\Delta L_{max}$,
in a disc with a flat rotation curve is proportional \kch{to} $R\,
|\Phi_s|^{1/2}_{CR}$.  \cite{DW15} predict that the maximum possible
change in orbital angular momentum for any individual star in a
trapped orbit ($\Delta L_{max}$) is set by the \rw{width} of the
capture region \citep[][their \S2.2.1]{DW15}.
\kch{Figure~\ref{fig:CapRegSize} illustrates t}he \kch{dependence of the maximum} width of the capture region ($R_{max}-R_{min}$) \kch{on} the fractional surface density for the spiral pattern ($\epsilon_\Sigma$) and its radius of corotation ($R_{CR}$). \rw{We expect that the maximum change in angular momentum per unit mass will be}  $\Delta L_{max}\approx v_c\,(R_{max}-R_{min})$.
\kch{W}hatever the functional form, \kch{t}he maximum possible change in angular momentum for a single star is of interest only in the sense that it is a limiting case, whereas the maximum amplitude for $\langle(\Delta L_z)^2\rangle^{1/2}$ depends on the \textit{distribution} of all the individual angular momenta of stars in trapped orbits.  \kch{Indeed,} \cite{Grand16} find a positive correlation between the amplitude of a perturbation and the degree of induced radial \kch{migration} in a suite of simulated \kch{disc} galaxies.

\cite{FBP16} predict that radial \kch{migration} will be more
important, for a given radial velocity dispersion $\sigma_R$, when the
value for $Q$ (equation~\ref{eqn:Q}) at corotation approaches unity
\kch{since self-gravity strongly enhances perturbations to the
potential at corotation.  They develop a formalism which they use to
compute a diffusion tensor in a razor thin \kch{disc} and find that
diffusion is parallel to the angular momentum axis at corotation,
consistent with radial migration.  In the series of models we use for}
Figure~\ref{fig:RMEII_CaptureParameters_CR} we hold the radial
velocity dispersion profile constant and vary the radius of
corotation.  \rw{The radial dependence for Q(R) would then be given by 
$Q \propto (R
\,e^{-R/R_d})^{-1}$} and \kch{would have} a minimum value of $Q\sim
0.7$ at $R=R_d$.  \kch{Our models} do not sample values of corotation
inside $R=4$~kpc, but measures of the integrated fraction with
unconstrained radial range of evaluation increase steadily toward the
galactic centre.  \kch{In
Figure~\ref{fig:RMEII_CaptureParameters_CR},} $Q = 1$ at $R =
5.5$~kpc, \kch{corresponding to a more steep decrease in the values
for the integrated fraction at} $R\gtrsim 5.5$~kpc.  \kch{The
comparison between \citeauthor{FBP16}'s results and ours is not
direct. \citeauthor{FBP16} argue that a lower velocity dispersion
leads to a stronger perturbation which stimulates more migration and
we argue that a stronger perturbation traps more orbits, but the
integrated fraction is lower for higher velocity dispersion.}
Additionally, our Model~Q, which specifically sets the radial velocity
dispersion so that $Q=1.5$ at all radii, \kch{has the lowest values
for radial velocity dispersion profile at all radii (illustrated in
Figure~\ref{fig:RMEII_sigR_exp_vs_Q}) and} the highest values for the
integrated fraction at all corotation radii.  This is a \kch{somewhat}
obvious result \kch{and} would presumably be more evident for lower
values for $Q$.

\kref{Spiral patterns that corotate or nearly corotate with the disc can induce radial migration over large radial distances \citep[][are examples of this phenomenon in N-body simulation]{WBS11,BSW13,DVH13,Grand16}.  
In \cite{DW15} \S4, we discuss how the radial range of the capture region depends on the radial rate of divergence between the spiral patten speed and the circular orbital frequency for stars; this divergence rate equals zero for corotating spiral patterns and thus the capture region spans the full radial range of the spiral arm.
The underlying assumptions for our models in this study have a divergence rate set by a flat rotation curve and radially constant pattern speed.  We have not explored other divergence rates, but we expect the integrated fraction to decrease with increasing kinematic temperature in all cases since the underlying physics will be the same no matter the size of the capture region.
A complication we have not addressed in this work is the consequence of overlapping resonances from multiple patterns \citep[e.g.,][]{MF10} or multiple transient modes \citep{SC14}.
}

\section{Conclusions}\label{sec:Conclusions}

This is the second of a series of papers seeking to \rwqq{quantify the
physical conditions under which  radial
\kch{migration} is an important process in the evolution of} \kch{disc} galaxies.  The efficiency of radial
\kch{migration} \rwqq{is directly related to the fraction of \kch{disc} stars
captured in trapped orbits, which is investigated here}.  We here apply the capture criterion \rwqq{we obtained in an earlier paper (a} brief review is given in
Appendix~\ref{sec:Capture} of this work) to a series of \kch{models}
of a spiral galaxy (described in \kch{\S\ref{sec:Approach}}) in order
to investigate how \kch{the adopted models and parameter values
affect} the initial distribution of the fraction of stars in trapped
orbits ($\mathcal{F}_{R}(R)$) and the integrated fraction
($\mathcal{F}_{\Delta R}$) over one of several specified radial ranges
($\Delta R$).

For any given spiral pattern, $\mathcal{F}_{R}(R)$ depends on the
\kch{distribution of orbital angular momentum} near corotation
($R_{CR}$) \kch{since whether \rwqq{or not} a star is in a trapped
orbit is well approximated by whether \rwqq{or not} its guiding centre
radius is within the capture region}.  At any given radial coordinate
$R$, the velocity dispersion characterises the width of the azimuthal
velocity distribution, and thus the distribution of orbital angular
momentum \rw{at} that radius.  The capture criterion specifies
\kch{the} range of orbital angular momenta \kch{at a given spatial
coordinate} for a star to be in a trapped orbit.  Therefore the
fraction of \kch{a} population that meets the capture criterion at
radius $R$ depends on the local velocity dispersion.  \kch{We use the
radial velocity dispersion as a proxy for the distribution of angular
momentum since it is \rw{more easily observed and quantified}.}  The
profile for $\mathcal{F}_{R}(R)$ for populations with high velocity
dispersion has a radial range that is greater than the size of the
capture region since the radial \rwqq{epicyclic} excursions (\rwqq{parameterised by the maximum amplitude, $X$)} of these stars from
their guiding centre radii (\rwqq{$R_L$, related to orbital} angular momentum by
equation~\ref{eqn:RL}) scales with the radial velocity dispersion (by
equation~\ref{eqn:X}).  The peak of \kch{the distribution of the
fraction of stars in trapped orbits has a} smaller \kch{amplitude} for
populations with larger velocity dispersion.  The offset of \kch{the}
peak of the distribution of stars in trapped orbits from the radius of
corotation \kch{($R_{peak}-R_{CR}$)} to larger radii is greater for
populations with higher velocity dispersion
(equation~\ref{eqn:RPeakPredict}) due to the slower mean orbital
velocity associated with asymmetric drift
(\S\ref{sec:RadialDistributionSig}).

\kch{T}he integrated fraction ($\mathcal{F}_{\Delta R}$) \kch{is the
measure we use} to quantify how the fraction \rw{of} stars in trapped
obits \kch{depends} on the radial velocity dispersion.  We evaluate
$\mathcal{F}_{\Delta R}$ \rw{by integrating the distribution of the trapped fraction over several specified  radial ranges} \kch{($\Delta
R$), where we use two classification schemes for $\Delta R$.  These
are the radial ranges that have either fixed radial width centred
around corotation} ($\Delta R_1$ and $\Delta R_2$) or an unconstrained
radial range ($\Delta R_{5\%}$, $\Delta R_{25\%}$, and $\Delta
R_{FWHM}$) \kch{defined by the range} within which the value for
$\mathcal{F}_{R}$ is greater than \kch{$5\%$, $25\%$, and half the
maximum amplitude of the fraction of stars in trapped orbits \rw{as a function of radius}}.
\kch{We use the combination of these fixed and unconstrained measures
for $\Delta R$ to gain the following insights into how the
distribution of stars in trapped orbits depends on the adopted model
and parameter values.}

\textit{\kch{T}he integrated fraction is smaller for stellar
populations with higher velocity dispersion}\kch{, consistent} with
the trend found by \cite{SSS12} (see \S\ref{sec:PreviousWork}).
\kd{In Paper~I, we analytically demonstrated that radial action, correlated with random orbital energy (i.e.,~the energy associated with non-circular motion), is mostly irrelevant to trapping.  Indeed, it is the azimuthal action, or angular momentum in the epicyclic limit, that is primarily responsible for trapping.  The kinematics of a \textit{population} of stars at \rwqq{2D} coordinate $\mathbf{x}$ can be described by the shape of a velocity ellipsoid having ratio $\sigma_R/\sigma_\phi$.  The spread in the orbital angular momentum distribution (related to $\sigma_\phi$ at $R$ through the individual angular momenta given by $L_z = R\,v_\phi$ per unit mass) is thus correlated with the radial velocity dispersion.  In Figure~\ref{fig:RMEII_VelocityDistribution}, we illustrate how the spread in orbital angular momentum and asymmetric drift determine the fraction of stars in trapped orbits at coordinate $\mathbf{x}$.}
\kch{Figure~\kd{\ref{fig:RMEII_FitSigma_AD}} shows a linear fit
(equation~\ref{eqn:CapturedFractionLinearFit}) for the integrated
fraction \kd{(total fraction of stars in trapped orbits)} as a function of radial velocity dispersion for each measure
of $\Delta R$ for \kd{the model that assumes} constant $\sigma_R$.}
\kd{We argue that a similar trend exists for any reasonable 
distribution function since the decrease in the integrated fraction fundamentally
stems from the broadening of the distribution of angular momentum.}

The
fitting constants are given in Table~\ref{tbl:FitLine}, where the
maximum fraction of stars in trapped orbits over the radial range of
the capture region ($\mathcal{F}_{5\%}$ as $\sigma_R\rightarrow 0$) is
well approximated by $\mathcal{F}_{\Delta 5\%, max}=2/\pi\sim64\%$
(\S\ref{sec:RadialDistributionSig}).  Further, we investigated the
effect of using a radially dependent radial velocity dispersion
profile (\S\ref{sec:CapturedFractionSigRadDep}) and found that there
are some \rwqq{differences} in the distribution of stars in trapped orbits,
but that the differences in the integrated fraction are not
particularly significant as long as the radial velocity dispersion
profile has a shallow radial gradient over \rwqq{the radial range of interest -  as} is the case
for all models \rwqq{in this paper.}

\kch{The \rwqq{distribution of stars (initially)} in trapped orbits
for populations with h}igh velocity dispersion \kch{has a large radial
range, spanning} \rw{of} order a few \kch{disc} scale lengths.  This
is a consequence of the large radial excursions from the guiding
centre \rwqq{radius for stars that meet the capture criterion \kch{(that the guiding centre radius lie} 
within the capture region)}.  However, the fraction of stars that meet
the capture criterion at large radial distances from $R_{CR}$ is
small.

The maximum value for the integrated fraction $\mathcal{F}_{\Delta R}$ over a given radial range ($\Delta R$) is determined by the amplitude of the spiral perturbation at corotation.

\kd{The integrated fraction ($\mathcal{F}_{\Delta R}$) for the fiducial Models~$\Sigma$, Q, and W},  is nearly \kd{independent} ($\pm \sim 10\%$) \kd{of corotation radius} ($R_{CR}$) \kd{or patten speed ($\Omega_p = v_c/R_{CR}$).  Each model assumes} an appropriate radially dependent radial velocity dispersion \kd{profile ($\sigma_R(R)$).}  The mean value for $\mathcal{F}_{\Delta R}$ in these models depends on the normalisation of the radial velocity dispersion profile \kd{and the assumed amplitude for the 
spiral pattern.}

The RMS change in orbital angular momentum, $\langle (\Delta
L_z)^2\rangle^{1/2}$, around the radius of corotation of a spiral
pattern depends on the fraction of stars in trapped orbits, which we
have explored in this paper.  $\langle (\Delta L_z)^2\rangle^{1/2}$
also depends on the amplitude of the oscillations in orbital angular
momentum for stars in trapped orbits (where the maximum amplitude
oscillation for a guiding centre radius is given by the size of the
capture region) and the time-scale for maximising $\langle (\Delta
L_z)^2\rangle^{1/2}$ due the these oscillations.  Should the spiral
disrupt on the time-scale for which each of these ingredients is
maximal, $\langle (\Delta L_z)^2\rangle^{1/2}$ would be maximised and
radial \kch{migration} would be most efficient for a single spiral.
In \S\ref{sec:UpperLimit}, we argue that time dependent processes are
an important part of understanding the efficiency \kch{of} radial
\kch{migration} from a single spiral pattern, and thus $\langle
(\Delta L_z)^2\rangle^{1/2}$, and the \textit{initial} fraction of
stars in trapped orbits should be treated as an upper limit.  We \rwqq{defer 
exploration of the importance of scattering processes, the
\textit{distribution} of orbital angular momentum oscillation
amplitudes, the \textit{distribution} of time-scales for trapped
orbits, and the consequences of a time-dependent spiral amplitude to} 
later papers in this series.

\section*{Acknowledgments}
It is a pleasure to thank \kd{J.~Sellwood, R.~\rwqq{Sch\"onrich} and P.~McMillan} for helpful discussions during the course of this \kref{study.
We also thank the anonymous referee for thoughtful comments that have improved the quality of this work.}
This material is based upon work supported by the \kch{US} National Science Foundation Graduate Research Fellowship under Grant No.~DGE-1232825, National Science Foundation Grants AST-0908326 and OIA-1124403, and an American Association of University Women Dissertation Fellowship.  \kch{RFGW thanks the \rw{Leverhulme  Trust for a Visiting Professorship, held at the University of Edinburgh. \kd{She is also grateful for support from her sister, Katherine Barber.} This work was performed in part at Aspen Center for Physics, which is supported by National Science Foundation grant PHY-1607611. RFGW thanks all at ACP for the wonderful environment for physics they have created and maintained}. }

\bibliographystyle{mn2e} 
\bibliography{mybibliography}

\appendix
\section{\rw{The} Capture Criterion}\label{sec:Capture}

In Paper~I, we derived a \lq\lq capture criterion" that can be used to
determine whether or not a \rw{star in an infinitely thin (2D) disc} is in a trapped orbit.  As
defined in \S\ref{sec:Introduction}, trapped orbits occur near the
radius of corotation with a spiral pattern (or any density
perturbation) and are characterised by having oscillatory orbital
angular momentum with little to no change in random orbital energy \rw{(as also discussed in \cite{SB02}).}  A
star that is in a trapped orbit could migrate radially should the
spiral pattern be transient.  In this section we briefly review the
capture criterion, but refer the reader to Paper~I for a \kch{more
complete} derivation and discussion.

The capture criterion can be applied to any \kch{disc} star given its 4D phase space coordinate ($R$,$\phi$,$v_R$,$v_\phi$), the amplitude of the spiral perturbation to the potential at the radius of corotation, $|\Phi_s|_{CR}$, and the slope of the rotation curve at corotation \citep[see][equation~32]{DW15}.

It is both informative and convenient to express the capture criterion in terms of orbital energy and orbital angular momentum.  The orbital energy in the inertial frame ($E$) and orbital angular momentum ($L_z$) can be combined to form an expression for a conserved quantity known as the Jacobi integral \citep[][eqns.~3.113]{BT08},
\begin{equation}\label{eqn:EJ}
E_J = E - \Omega_p L_z = \frac{1}{2}|\dot{\mathbf{x}}|^2 + \Phi_{eff}(\mathbf{x}) ,
\end{equation}
with
\begin{equation}\label{eqn:EffectivePotential}
\Phi_{eff}(\mathbf{x}) \equiv \Phi(\mathbf{x}) -\frac{1}{2}|\mathbf{\Omega_p} \times \mathbf{x}|^2
\end{equation}
being the effective potential \citep[][eqns.~3.114]{BT08}, and $\mathbf{x}$ and $\mathbf{\dot{x}}$ being, respectively, the position and velocity of the star in the rotating frame.  Since the spiral potential is time-independent in the rotating frame, it is in this frame that the Jacobi integral is conserved.  

\begin{figure}
\begin{center}
\includegraphics[scale=1]{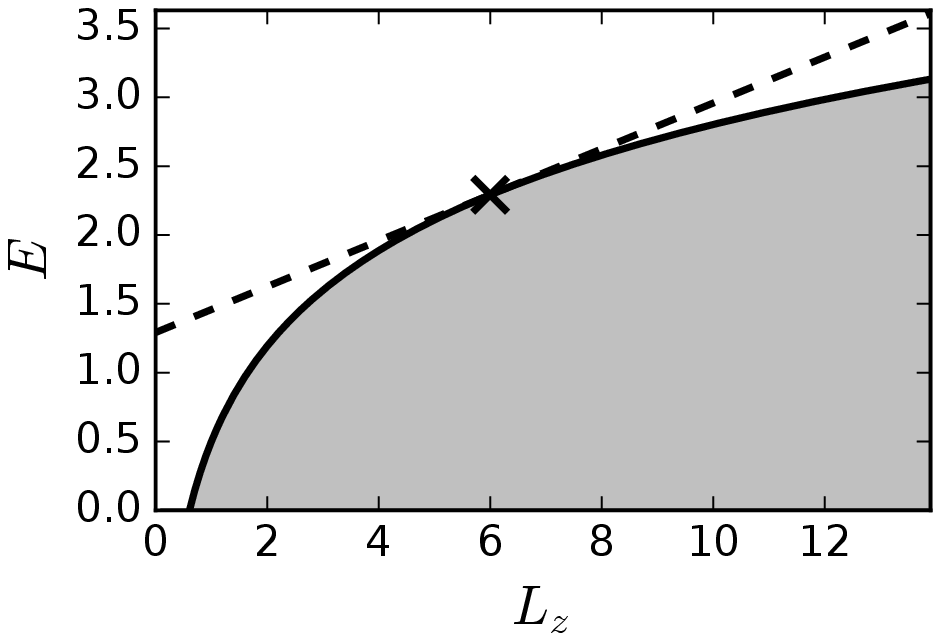}
\caption{Lindblad diagram illustrating how changes in orbital angular
momentum relate to changes in orbital energy via the Jacobi
integral. The solid curve represents circular orbits and the shaded
region is inaccessible since an orbit cannot have energy less than
circular.  Vertical distance from the solid curve represents random
orbital energy. The \lq X' marks corotation and the dashed line has
slope equal to the pattern speed ($\Omega_p$) of the spiral pattern.
The Jacobi integral is conserved along lines with slope $\Omega_p$, so
stars near corotation that have changes in orbital angular momentum
will have little to no change in random orbital energy.  This diagram
uses units where $v_c= L_z/R = 1$ and is \rw{only slightly modified from
Figure~1 from \citet{SB02}}.}
\label{fig:LindbladDiagram}
\end{center}
\end{figure}

Figure~\ref{fig:LindbladDiagram} shows a Lindblad diagram for stars in a \kch{disc} with a flat rotation curve.  The Lindblad diagram illustrates the relationship between the Jacobi integral and the conservation of orbital circularity at corotation.  The time derivative of equation~\ref{eqn:EJ} shows that changes in a star's orbital angular momentum are linearly related to changes in its orbital energy with slope equal to the spiral pattern speed (represented by a dashed line in Figure~\ref{fig:LindbladDiagram}).  In a \kch{disc} with a flat rotation curve, stars in circular orbits (solid curve) have logarithmically increasing orbital energy with linearly increasing orbital angular momentum, where vertical distance from this curve represents random orbital energy.  At corotation (shown as an \lq X' in Figure~\ref{fig:LindbladDiagram}), the slope of the line representing pattern speed is parallel to slope of constant orbital circularity.  Since the Jacobi integral ($E_J$) is conserved along lines with slope equal to the pattern speed ($\Omega_p$), stars in trapped orbits do not have significant changes in orbital eccentricity with changes in orbital angular momentum around corotation.

The generalized capture criterion is derived using action variables, but in order to write an expression in terms of orbital energy and angular momentum we assume the epicyclic approximation.  This approximation describes mildly non-circular orbits as simple harmonic oscillations about a circularly orbiting guiding centre.
The epicyclic frequency of radial oscillations is given by \citep[][equation~3.80]{BT08},
\begin{equation}\label{eqn:kappa}
\kappa^2(R_L)=\left( R\dfrac{d\Omega_c^2}{dR}+4\Omega_c^2\right)_{R_L} ,
\end{equation}
where $\Omega_c$ is the circular orbital frequency, and the mean orbital radius\footnote{Note that this approximation is only strictly true for SHM, requiring radial oscillations be within a range where the surface density is approximately constant.} (i.e.~\lq\lq guiding centre radius") is
\begin{equation}\label{eqn:RL}
R_L = R\,\dfrac{v_\phi}{v_c}.
\end{equation}
The random orbital energy in the inertial frame is
\begin{equation}\label{eqn:Eran}
E_{ran}=E-E_c(R_L),
\end{equation}
where $E_c(R)$ is the energy associated with a circular orbit in the underlying axisymmetric potential at radius $R$.

The form of the capture criterion appropriate for our assumed model is \citep[][their equation~34]{DW15}
\begin{equation}\label{eqn:CaptureCriterion}
-1 < \Lambda_{nc,2}(t) \leq 1 ,
\end{equation}
where \citep[][from their equations~11 \& 33]{DW15}
\begin{equation}\label{eqn:Lambda_nc2}
\Lambda_{nc,2}(t) \equiv \dfrac{E_J\kch{(R,\phi,v_R,v_\phi)}-h_{CR}}{|\Phi_s|_{CR}} - \left(\dfrac{R_L(\kch{R,v_\phi,}t)}{R_{CR}}\right) \left( \dfrac{E_{ran}\kch{(R,\phi,v_R,v_\phi,t)}}{|\Phi_s|_{CR}}\right)
\end{equation}
and $h_{CR}$ is the Jacobi integral (equation~\ref{eqn:EJ}) for a star at corotation in the underlying axisymmetric potential ($\Phi_0(R_{CR})$).

\label{lastpage}

\end{document}